\documentclass[a4paper,12pt]{article}

\usepackage{mathrsfs}
\usepackage{graphicx}
\usepackage{bm}
\usepackage{amssymb}
\usepackage{amsmath}
\usepackage{ascmac}
\usepackage{array}

\usepackage{cite}

\begin{document}



\title{Information flow and entropy production on Bayesian networks}
\author{Sosuke Ito\footnote{\scriptsize Department of Physics, Tokyo Institute of Technology, Oh-okayama 2-12-1, Meguro-ku, Tokyo, 152-8551, Japan} \ and Takahiro Sagawa\footnote{\scriptsize Department of Applied Physics, The University of Tokyo, 7-3-1 Hongo, Bunkyo-ku, Tokyo 113-8656, Japan}}

\date{}

\maketitle


In this article, we review a general theoretical framework of thermodynamics of information on the basis of Bayesian networks.
This framework can describe a broad class of nonequilibrium dynamics of multiple interacting systems with complex information exchanges. 
For such situations, we discuss a generalization of the second law of thermodynamics including information contents.
A key concept here is an informational quantity called the transfer entropy, which describes the directional information transfer in stochastic dynamics.
The generalized second law gives the fundamental lower bound of the entropy production in nonequilibrium dynamics, and sheds modern light on the paradox of ``Maxwell's demon'' that performs measurements and feedback control at the level of thermal fluctuations.

\tableofcontents

\section{Introduction\label{Sec_Introduction}}

\subsection{Background\label{Sec_Background}}

The second law of thermodynamics is one of the most fundamental laws in physics, which identifies the upper bound of the efficiency of heat engines~\cite{Callen}.  The second law has been established in the nineteenth century, after numerous failed trials to invent a perpetual motion of the second kind.  Today we realize that it is not possible; one can never extract a positive amount of work from a single heat bath in a cyclic way, or equivalently, the entropy of the whole universe never decreases.

While thermodynamics has been formulated for macroscopic systems, thermodynamics of small systems has been developed over the last two decades.
Imagine a single Brownian particle in water.  The particle goes to thermal equilibrium in the absence of external driving, because water plays the role of a huge heat bath.  In this case, even a single small particle can behave as a thermodynamic system.  Moreover, if we drive the particle by applying a time-dependent external force, the particle goes far from equilibrium.
Such a small stochastic system is an interesting playing field to investigate ``stochastic thermodynamics''~\cite{Sekimoto, Seifert}, which is a generalization of thermodynamics by including the role of thermal fluctuations explicitly.
We can show that, in small systems, the second law of thermodynamics can be violated stochastically, but is never violated on average.
The probability of the violation of the second law can quantitatively be characterized by the fluctuation theorem~\cite{Cohen,Gallavotti,EvansF,Jarzynski1,Jarzynski2,Seifert2005}, which is a prominent discovery in stochastic thermodynamics.
From the fluctuation theorem, we can reproduce the second law of thermodynamics on average.
Stochastic thermodynamics is applicable not only to a simple Brownian particle~\cite{Wang}, but also to much more complex systems such as RNA foldings~\cite{Liphardt,Collin} and biological molecular motors~\cite{Hayashi}.

More recently, stochastic thermodynamics has been extended to information processing processes~\cite{review}.
The central idea is that one can utilize the information about thermal fluctuations to control small thermodynamic systems.
Such an idea dates back to the thought experiment of ``Maxwell's demon'' in the nineteenth century~\cite{Maxwell}.
The demon can perform a measurement of the position and the velocity of each molecule, and manipulate it by utilizing the obtained measurement outcome.
By doing so, the demon can apparently violate the second law of thermodynamics, by adiabatically decreasing the entropy.
The demon has puzzled many physicist over a century~\cite{Demon,Szilard, Brillouin, Bennett, Landauer}, and it is now understood that the key to understand the consistency between the demon and the second law is the concept of information~\cite{Shannon,Cover-Thomas,Schreiber}, and that  the demon can be regarded as a feedback controller.

The recent  theoretical progress in this field has led to a unified theory of information and thermodynamics, which may be called information thermodynamics~\cite{review,Touchette1,Touchette2,CaoFeito,Sagawa1,Ponmurugan,Fujitani,HorowitzVaikun,Broeck,HorowitzParro,Ito1,AbreuSeifert,Sagawa2012,Sagawa2,Kundu,StillCrooks,Sagawa3}.  The thermodynamic quantities and information contents are treated on an equal footing in information thermodynamics.
In particular, the second law of thermodynamics has been generalized by including an informational quantity called the mutual information.
The demon is now regarded as a special setup in the general framework of information thermodynamics.
The entropy of the whole universe does not decrease even in the presence of the demon, if we take into account the mutual information as a part of the total entropy.
Information thermodynamics  has recently been experimentally studied with a colloidal particle~\cite{Toyabe,Berut,Berut2,Roldan} and a single electron~\cite{Koski}.  

Furthermore, the general theory of information thermodynamics is not restricted to the conventional setup of Maxwell's demon, but is applicable to a variety of dynamics with complex information exchanges.
In particular, information thermodynamics is applicable to autonomous information processing~\cite{Kim,MunakataRosinberg1,Munakata,MandalJarzynski,BaratoSeifert0,Strasberg,HSP,Ito,Allahverdyan,HartichSeifert,HorowitzEsposito,HorowitzSandberg,Naoto,Shiraish2}, and is further applicable to sensory networks and biochemical signal transduction~\cite{BaratoSeifert,Bo,HartichSeifert2,SartoriHorowitz,Ito2}.
Such complex and autonomous information processing can be formulated in a unified way on the basis of Bayesian networks~\cite{Ito}; this is the main topic of this chapter.
An informational quantity called the transfer entropy~\cite{Schreiber}, which represents the directional information transfer, is shown to play a significant role in  the generalized second law  of thermodynamics on Bayesian networks.

\subsection{Basic ideas of information thermodynamics\label{Sec_Basic}}

\begin{figure}[htbp]
 \begin{center}
  \includegraphics[width=110mm]{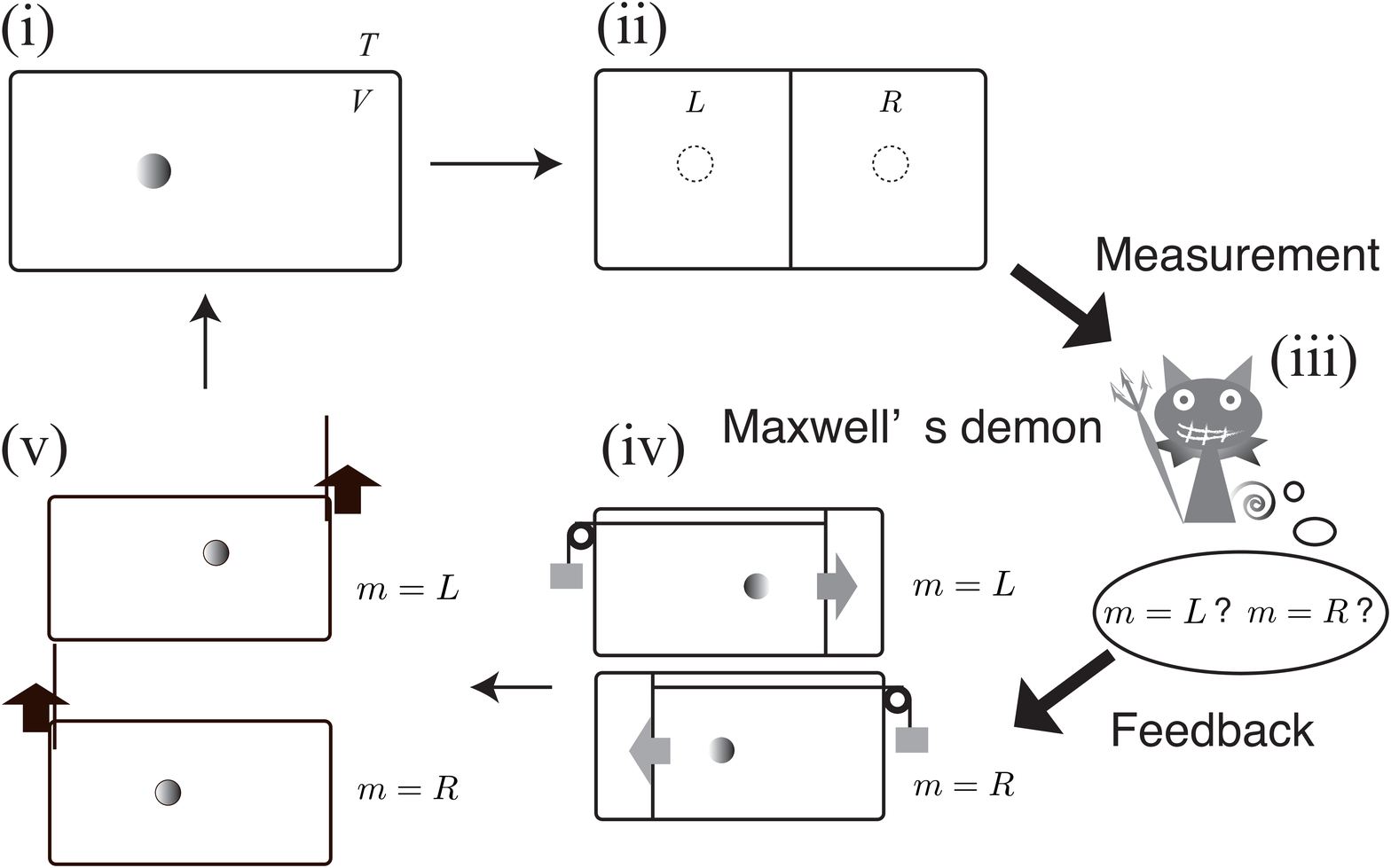}
 \end{center}
 \caption{Schematic of the Szilard engine. The demon obtains measurement outcome $m=L$ (left) or $m=R$ (right), corresponding to one bit ($=\ln 2$) of information.  The demon then extracts $k_{\rm B}T \ln 2$ of work by feedback control.} 
 \label{Szilard}
\end{figure}

Before proceeding to the main part of this chapter, we briefly sketch the basic idea of information thermodynamics.
The simplest model of Maxwell's demon is known as the Szilard engine~\cite{Szilard}, which is shown in  Fig.~\ref{Szilard}.
We consider a single particle in a box with volume $V$ that is in contact with a heat bath at temperature $T$.
The time evolution of the Szilard engine is as follows.
(i) The particle is in thermal equilibrium, and the position of the particle is uniformly distributed. (ii) We divide the box by inserting a barrier at the center of the box.  (iii) The demon performs a measurement of the position of the particle, and finds it in the left or right box with probability $1/2$.  
The obtained information is one bit, or equivalently $\ln 2$ in the natural logarithm.
(iv) If the particle is found in the left (right) box, then the demon slowly moves the barrier to the right (left) direction, which is  feedback control depending on the measurement outcome.  This process is assumed to be isothermal and quasi-static.
(v)  The partition is removed, and the particle returns to the initial equilibrium state.  

In step (iv), the single-particle gas is isothermally expanded and a positive amount of work is extracted.  The amount of the work  can be calculated by using the equation of states of the single-particle ideal gas (i.e., $pV = k_{\rm B}T$ with $k_{\rm B}$ the Boltzmann constant):
\begin{equation}
\int_{V/2}^V \frac{k_{\rm B}T}{V'} dV' = k_{\rm B}T \ln 2.
\end{equation}
This is obviously positive, while the entire process seems to be cyclic.  
The crucial point here is that the extracted work $k_{\rm B}T \ln 2$ is proportional to the obtained information $\ln 2$, which suggests the fundamental information-thermodynamics link.

\subsection{Outline of this chapter\label{Sec_Outline}}

 In the following, we present an introduction to a theoretical framework of information thermodynamics  on the basis of Bayesian networks.  This chapter is organized as follows.
In Sec.~\ref{Sec_Brief}, we briefly review the basic properties of information contents: the Shannon entropy, the relative entropy, the mutual information, and the transfer entropy.
In Sec.~\ref{Sec_Stochastic}, we review stochastic thermodynamics by focusing on a simple case of Markovian dynamics.  In particular, we discuss the concept of entropy production.
In Sec.~\ref{Sec_Bayesian}, we review the basic concepts and terminologies of Bayesian networks.
In Sec.~\ref{Sec_Information1}, we discuss the general theory of information thermodynamics on Bayesian networks, and derive the generalized second law of thermodynamics including the transfer entropy.
In Sec.~\ref{Sec_Examples}, we apply the general theory to special situations such as repeated measurements and feedback control.
In particular, we discuss the relationship between our approach based on the transfer entropy and another approach based on the dynamic information flow~\cite{Allahverdyan,HartichSeifert,HorowitzEsposito,HorowitzSandberg,Naoto,Shiraish2}.
In Sec.~\ref{Sec_Summary}, we summarize this chapter, and discuss the future prospects of information thermodynamics.

\section{Brief review of information contents\label{Sec_Brief}}

In this section, we review the basic properties of several informational quantities. We first discuss various types of entropy: the Shannon entropy, the relative entropy, and the mutual information~\cite{Shannon, Cover-Thomas}.  We next discuss the transfer entropy that quantifies the directional information transfer~\cite{Schreiber}.

\subsection{Shannon entropy\label{Sec_Shannon}}
We first discuss the Shannon entropy, which characterizes the randomness of probability variables. Let $x$ be a probability variable with probability distribution $p(x)$.
We first define a quantity called the stochastic Shannon entropy:
\begin{equation}
s(x):=- \ln p(x),
\end{equation}
which is large if $p(x)$ is small.
The ensemble average of $s(x)$ over all $x$ is equal to the Shannon entropy:
\begin{align}
\langle s(x) \rangle := - \sum_{x} p(x) \ln p(x).
\label{Shannon1}
\end{align}
We note that $\langle \cdots \rangle$ describes the ensemble average throughout this paper.
Since $0 \leq p(x)  \leq 1$, we have $s(x) \geq 0$, and therefore 
\begin{equation}
\langle s(x) \rangle \geq 0.
\end{equation}

Let $y$ be another probability variable that has the joint probability distribution with $x$ as $p(x,y)$.  The conditional probability of $x$ under the condition of $y$ is given by $p(x|y) := p(x,y) / p(y)$, which is the Bayes rule.   We define the stochastic conditional Shannon entropy  as
\begin{equation}
 s(x|y) :=- \ln p(x|y),
\end{equation}
whose ensemble average is the conditional Shannon entropy:
\begin{align}
\langle s(x|y) \rangle = -\sum_{x, y} p(x, y) \ln p(x|y).
\label{condShannon}
\end{align}

\subsection{Relative entropy\label{Sec_Relative}}

We next introduce the relative entropy (or the Kullback-Leibler divergence), which is a measure of the difference of two probability distributions.  We consider two probability distributions $p$ and $q$ on the same probability variable $x$. 
The relative entropy  between the probability distributions  is defined as
\begin{align}
D_{\rm KL} (p\| q)  := \sum_{x} p(x) \ln \frac{p(x)}{ q(x) } =  \sum_{x} p(x) [\ln p(x) - \ln q(x) ].
\end{align}
By introducing the stochastic relative entropy as
\begin{equation}
d_{\rm KL}(p(x)\| q(x)) :=  \ln p(x) - \ln q(x),
\label{stochastic_relative}
\end{equation}
we write the relative entropy as
\begin{equation}
D_{\rm KL} (p\| q) = \langle d_{\rm KL}  (p(x)\| q(x))  \rangle.
\end{equation}

The relative entropy is always nonnegative. To show this, we use the Jensen inequality~\cite{Cover-Thomas}
\begin{equation}
\langle -\ln [q(x)/p(x)] \rangle \geq -\ln \langle q (x)/ p(x)\rangle,
\end{equation}
which is a consequence of the concavity of the logarithmic function. We then have
\begin{equation}
\begin{split}
D_{\rm KL} (p\| q) &\geq -\ln \left\langle \frac{q (x)}{p(x)} \right\rangle\\
&= -\ln \sum_x p(x) \frac{q (x)}{p(x)} \\
&= -\ln \sum_x q (x) \\
&= 0,
\end{split}
\label{nonnegativerelative}
\end{equation}
where we used $\sum_x q(x) =1$.
We note that $D_{\rm KL} (p(x)\| q(x))  = 0$ if and only if $q(x) = p(x)$.

We can also show the nonnegativity of the relative entropy in a slightly different way as follows.
We first note that
\begin{equation}
\langle e^{-d_{\rm KL}  (p(x)\| q(x))} \rangle= 1,
\label{relative_FT}
\end{equation}
because
\begin{equation}
\langle e^{-d_{\rm KL}  (p(x)\| q(x))} \rangle = \left\langle \frac{q(x)}{p(x)} \right\rangle = \sum_{x}p(x)\frac{q(x)}{p(x)} = \sum_{x}q(x) = 1.
\end{equation}
By applying the Jensen inequality to the exponential function that is convex, we have
\begin{align}
\langle \exp (-d_{\rm KL}  (p(x)\| q(x)) ) \rangle \geq \exp (- \langle  d_{\rm KL}  (p(x)\| q(x))  \rangle ).
\end{align}
Therefore, we obtain
\begin{equation}
1 \geq \exp (- D_{\rm KL} (p\| q)  ),
\end{equation}
which implies the nonnegativity of the relative entropy.  
We note that this proof is closely related to the fluctuation theorem as shown in Sec.~\ref{Sec_Stochastic}.

\subsection{Mutual information\label{Sec_Mutual}}

We discuss the mutual information between two probability variables $x$ and $y$, which is an informational measure of correlation~\cite{Shannon, Cover-Thomas}. 
The stochastic mutual information between $x$ and $y$ is defined as
\begin{align}
I(x : y) := \ln \frac{p(x, y)}{p( x)p( y)}= \ln p(x, y) - \ln p(x) - \ln p( y),
\label{mutual1}
\end{align}
which can be rewritten as the stochastic relative entropy between $p(x, y)$ and $p(x)p( y)$:
\begin{equation}
I(x : y)  = d_{\rm KL} (p(x, y)\|  p(x)p( y)).
\end{equation}
Its ensemble average is the mutual information:
\begin{equation}
\langle I(x : y)  \rangle= \sum_{x,y}p(x, y)\ln \frac{p(x, y)}{p( x)p( y)} = \langle d_{\rm KL} (p(x, y)\|  p(x)p( y)) \rangle.
\end{equation}
From the nonnegativity of the relative entropy, we have 
\begin{equation}
\langle I(x : y) \rangle \geq 0.
\end{equation}
The  equality is achieved if and only if $x$ and $y$ are stochastically independent, i.e., $p(x, y) = p(x) p(y)$.

\begin{figure}[htbp]
 \begin{center}
  \includegraphics[width=50mm]{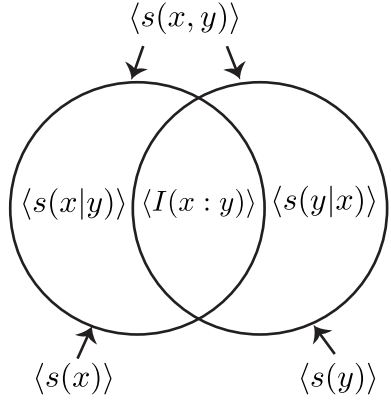}
 \end{center}
 \caption{Venn diagram of the relationship between the Shannon entropy and the mutual information.} 
 \label{Venn}
\end{figure}

The mutual information can also be rewritten as the difference of the Shannon entropy:
\begin{equation}
\begin{split}
\langle I(x : y) \rangle&= \langle s(x) \rangle + \langle s(y) \rangle -  \langle s(x, y) \rangle\\
&= \langle s (x) \rangle  - \langle s(x| y) \rangle \\
&=\langle s (y) \rangle -  \langle s(y |x) \rangle.
\end{split}
\label{mutuals}
\end{equation}
From the nonnegativity of the conditional Shannon entropy, we find that the mutual information is bounded by the Shannon entropy:
\begin{equation}
\langle I(x : y) \rangle  \leq \langle s(x) \rangle, \ \ \langle I(x : y) \rangle  \leq \langle s(y) \rangle.
\end{equation}
Figure~\ref{Venn} shows a Venn diagram that summarizes the relationship between the Shannon entropy and the mutual information.

We can also define the stochastic conditional mutual information between $x$ and $y$ under the condition of another probability variable $z$ as
\begin{align}
I(x : y| z) := \ln \frac{p(x, y| z)}{p(x| z)p( y| z)} = d_{\rm KL} (p(x, y| z)\|  p(x| z)p( y| z)).
\label{mutual2}
\end{align}
Its ensemble average is the conditional mutual information:
\begin{equation}
\langle I(x : y | z)  \rangle := \sum_{x,y,z} p(x,y,z) \ln \frac{p(x, y| z)}{p(x| z)p( y| z)}.
\end{equation}
We have $\langle I(x : y|z) \rangle \geq 0$, where the equality is achieved if and only if $x$ and $y$ are  conditionally independent, i.e., $p(x, y| z) = p(x| z) p(y| z)$.

\subsection{Transfer entropy\label{Sec_Transfer}}

The directional information transfer between two stochastic systems can be characterized by an informational quantity called the transfer entropy~\cite{Schreiber}.  We consider a sequence  of two probability variables: $(x_1, x_2, \cdots, x_N, y_1, y_2, \cdots, y_N)$.
Intuitively, the states of interacting two systems $X$ and $Y$ at time $k$ ($=1,2, \cdots, N$) is given by $(x_k, y_k)$.
The time evolution of the composite system is characterized by the transition probability $p(x_{k+1}, y_{k+1} | x_1, y_1, \cdots, x_k, y_k)$, which is the probability of $(x_{k+1}, y_{k+1})$ under the condition of $( x_1, y_1, \cdots, x_k, y_k)$.  The joint probability of all the variables is given by
\begin{equation}
p(x_1, \cdots, x_N, y_1,  \cdots, y_N) = \prod_{k=1}^{N-1}p(x_{k+1}, y_{k+1} | x_1, y_1, \cdots, x_k, y_k) \cdot p(x_1, y_1).
\end{equation}

We now consider the information transfer from system $X$ to $Y$ during time $k$ and $k+1$.  We define the stochastic transfer entropy as the stochastic conditional mutual information: 
\begin{equation}
\begin{split}
I_{k+1}^{\rm tr} (X\to Y) &:= I((x_1, \cdots, x_k) : y_{k+1} | y_1, \cdots, y_k) \\
&:= \ln \frac{p(x_1, \cdots, x_k, y_{k+1} | y_1, \cdots, y_k)}{p(x_1, \cdots, x_k  | y_1, \cdots, y_k)p(y_{k+1}  | y_1, \cdots, y_k)}.
\end{split}
\label{transfer_T}
\end{equation}
Its ensemble average is the transfer entropy:
\begin{equation}
\begin{split}
&\langle I_{k+1}^{\rm tr} (X\to Y) \rangle  \\
&:= \langle I((x_1, \cdots, x_k): y_{k+1} | y_1, \cdots, y_k) \rangle \\
&:= \sum_{x_1, \cdots, x_k, y_1, \cdots, y_{k+1}} p(x_1, \cdots, x_k, y_1,  \cdots, y_k, y_{k+1})  \ln \frac{p(x_1, \cdots, x_k,y_{k+1} | y_1, \cdots, y_k)}{p(x_1, \cdots, x_k  | y_1, \cdots, y_k)p(y_{k+1}  | y_1, \cdots, y_k)},
\end{split}
\end{equation} 
which represents the  information about past trajectory $(x_1, x_2, \cdots, x_k)$ of system $X$, which is newly obtained by system $Y$ from time $k$ to $k+1$.  
While the mutual information is symmetric between two variables in general, the transfer entropy is asymmetric between two systems $X$ and $Y$, as the transfer entropy represents the directional transfer of information.

Equality~(\ref{transfer_T}) can be rewritten as 
\begin{equation}
I_{k+1}^{\rm tr} (X\to Y) = I((x_1, \cdots, x_k): (y_1,  \cdots, y_k, y_{k+1})) - I((x_1, \cdots, x_k): (  y_1, \cdots, y_k)),
\label{transfer_T2}
\end{equation}
because
\begin{equation}
\begin{split}
&I((x_1, \cdots,x_k): y_{k+1} | y_1, \cdots, y_k) \\
&= \ln \frac{p(x_1, \cdots, x_k, y_1, \cdots, y_k, y_{k+1})p(y_1, \cdots, y_k)}{p(x_1, \cdots, x_k, y_1, \cdots, y_k)p(y_1, \cdots, y_k, y_{k+1})} \\
&= \ln \frac{p(x_1, \cdots, x_k,y_1,  \cdots, y_k, y_{k+1})}{p(x_1, \cdots, x_k)p( y_1, \cdots, y_k, y_{k+1})} - \ln \frac{p(x_1, \cdots,x_k, y_1, \cdots, y_k)}{p(x_1, \cdots, x_k)p(y_1, \cdots, y_k)} \\
&=  I((x_1, \cdots, x_k):  (y_1, \cdots, y_k,y_{k+1})) - I((x_1, \cdots, x_k): (  y_1, \cdots, y_k)).
\end{split}
\end{equation}
Equality~(\ref{transfer_T2}) clearly shows the meaning of the transfer entropy: the information about $X$ newly obtained  by $Y$.
We note that Eq.~(\ref{transfer_T}) can also be rewritten by using the stochastic conditional Shannon entropy:
\begin{equation}
I_{k+1}^{\rm tr} (X\to Y) = s(y_{k+1} |y_1, \cdots, y_k) - s(y_{k+1} | x_1, \cdots, x_k, y_1, \cdots, y_k).
\label{transfer_conditoinal_Shannon}
\end{equation}
Therefore, $\langle I_{k+1}^{\rm tr} (X\to Y) \rangle$ describes the reduction of the conditional Shannon entropy of $y_{k+1}$ due to the information gain about system $X$, which again confirms the meaning of the transfer entropy.

\section{Stochastic thermodynamics for Markovian dynamics\label{Sec_Stochastic}}

We review stochastic thermodynamics of Markovian dynamics~\cite{Sekimoto, Seifert}, which is a theoretical framework to describe  thermodynamic quantities such as the work, the heat and the entropy production, at the level of thermal fluctuations.  In particular, we discuss the second law of thermodynamics and the fluctuation theorem~\cite{Cohen,Gallavotti,EvansF,Jarzynski1,Jarzynski2,Seifert2005}. 


\subsection{Setup\label{Sec_Setup1}}

We consider system $X$ that stochastically evolves.
We assume the physical situation that system $X$ is attached to a single heat bath at inverse temperature $\beta := (k_{\rm B}T)^{-1}$, and that system $X$  is driven by external control parameter  $\lambda$ that describes, for example, the volume of the gas.
We also assume that  nonconservative force is not applied to system $X$ for simplicity.
Moreover, we assume that system $X$ does not include any odd variable that changes its sign with the time-reversal transformation (e.g., momentum).
The generalization beyond these simplification is  straightforward.

Although real physical dynamics are continuous in time, our formulation in this chapter is discrete in time.  Therefore, we discretize time  as follows.
Suppose that the real stochastic dynamics of system $X$ is parameterized by continuous time $t$.
We then focus on the state of  system $X$ only at discrete time $t_k :=k \Delta t$ ($k=1,2, \cdots, N$), where $\Delta t$ is a finite time interval.  
In the following, we  refer to time $t_k$ just as ``time $k$.''
Let $x_k$ be the state of system $X$ at time $k$.

\begin{figure}[htbp]
 \begin{center}
  \includegraphics[width=70mm]{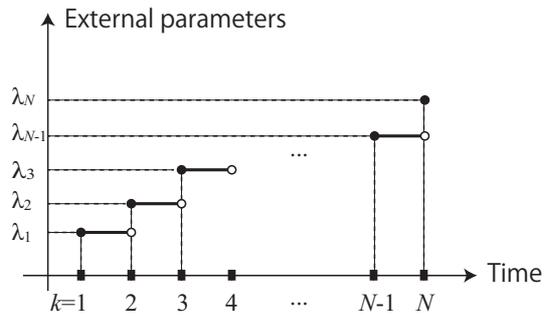}
 \end{center}
 \caption{The discretization of the time evolution of the external parameter.} 
 \label{heat_work}
\end{figure}

We next assume that $\lambda$ takes a fixed value $\lambda_k$ during time interval $t_k \leq t < t_{k+1}$.
The value of $\lambda$ is changed from $\lambda_k$ to $\lambda_{k+1}$ immediately before time $t_{k+1}$ (see also Fig.~\ref{heat_work}).
We here assume that the time evolution of $\lambda$ is predetermined independent of the state of $X$.

Let $p(x_k|x_{k-1}, \dots, x_1)$ be the conditional probability of state $x_k$ under the condition of past trajectory $x_1 \to \cdots \to x_{k-1}$.  
It is natural to assume that the conditional probability is determined by external parameter $\lambda_k$  that is fixed during time interval $t_k \leq t < t_{k+1}$; we can explicitly show the $\lambda_k$-dependence by writing $p(x_k|x_{k-1}, \dots, x_1; \lambda_k)$.

We also assume that the correlation time of the heat bath in the continuous-time dynamics is much shorter than $\Delta t$.
Under this assumption, the discretized time evolution $x_1 \to x_2 \to \cdots \to x_N$ can be regarded as Markovian.
We note that, if the continuous-time dynamics itself is Markovian, the discretized dynamics is obviously Markovian. 
From the Markovian assumption, we have
\begin{align}
p(x_k|x_{k-1}, \dots, x_1; \lambda_k )  = p(x_{k+1}|x_k; \lambda_k),
\label{notation_f}
\end{align}
which  we sometimes write as, for simplicity of notation, 
\begin{equation}
p(x_k|x_{k-1}) := p(x_{k+1}|x_k; \lambda_k).
\end{equation}
The joint probability distribution of $(x_1, x_2, \cdots, x_N)$ is then given by
\begin{eqnarray}
p(x_1, x_2, \cdots, x_N) := p(x_N|x_{N-1}) \cdots  p(x_3|x_2)p(x_2|x_1)p(x_1).
\label{Markov}
\end{eqnarray}
To make the notation simpler, we define  set $\mathcal{X} := \{ x_1, x_2, \cdots, x_N \}$, and denote
\begin{equation}
p(\mathcal{X}) := p(x_1, x_2, \cdots, x_N).
\end{equation} 
Strictly speaking,  set $\{ x_1, x_2, \cdots, x_N \}$ is not the same as vector $(x_1, x_2, \cdots, x_N)$.  However,  we sometimes do not distinguish them by notations for the sake of simplicity.


\subsection{Energetics\label{Sec_Energetics}}

We now consider the energy change in system $X$, and discuss the first  law of thermodynamics.
Let $E (x_k; \lambda_k)$ be the energy (or the Hamiltonian) of system $X$ at time $t_k$, which depends external parameter $\lambda_k$ as well as state $x_k$.
The energy change in system $X$ is decomposed into two parts: the heat and the work. 
The heat is the energy change in $X$ due to the stochastic change of the state of $X$ induced by the heat bath, and the work is the energy change due to the change of external parameter $\lambda$.
We stress that  the heat and the work are defined at the level of stochastic trajectories in stochastic thermodynamics~\cite{Sekimoto}.

The heat absorbed by system $X$ from the heat bath during time interval $t_k \leq t < t_{k+1}$ is given by
\begin{equation}
Q_k := E (x_{k+1}; \lambda_k ) - E (x_k; \lambda_k),
\label{heat}
\end{equation}
which is a stochastic quantity due to the stochasticity of $x_k$ and $x_{k+1}$.
On the other hand, the work is performed at time $k$ at which the external parameter is changed.  The work performed on system $X$ at time $k$ is given by  (see also Fig.~\ref{heat_work})
\begin{equation}
W_k  := E(x_k; \lambda_k) - E  (x_k; \lambda_{k-1}),
\label{work}
\end{equation}
which is also a stochastic quantity.

The total heat absorbed by system $X$ from time $1$ to $N$ along trajectory $(x_1, x_2, \cdots, x_N)$ is then given by
\begin{equation}
Q := \sum_{k=1}^{N-1}Q_k,
\end{equation}
and the total work is given by
\begin{equation}
W := \sum_{k=2}^{N}W_k. 
\end{equation} 
It is easy to check that the total heat and the work satisfy the first law of thermodynamics:
\begin{equation}
\Delta E = Q+W,
\label{first_law}
\end{equation}
where 
\begin{equation}
\Delta E := E (x_N, \lambda_N) - E (x_1, \lambda_1)
\end{equation}
is the total energy change.
We note that Eq.~(\ref{first_law}) is the first law at the level of individual trajectories.

\subsection{Entropy production and fluctuation theorem\label{Sec_Entropy1}}

We next consider the second law of thermodynamics.
We start from the concept of the detailed balance, which is satisfied in the absence of any nonconservative force.
The detailed balance is given by, from time $k$ to $k+1$,
\begin{equation}
p (x_{k+1} | x_k; \lambda_k) e^{-\beta E (x_k; \lambda_k)} = p_k (x_{k} | x_{k+1}; \lambda_k) e^{-\beta E (x_{k+1}; \lambda_k)},
\label{detailedbalance}
\end{equation}
where $p_k (x_{k} | x_{k+1}; \lambda_k)$ describes the ``backward'' transition probability from $x_{k+1}$ to $x_k$ under external parameter $\lambda_k$.
Equality~(\ref{detailedbalance}) can also be written as, from the definition of heat (\ref{heat}),
\begin{equation}
\frac{p (x_{k+1} | x_k; \lambda_k)}{p (x_{k} | x_{k+1}; \lambda_k)}   =   e^{-\beta Q_k }.
\label{detailedbalance1}
\end{equation}
The detailed balance condition (\ref{detailedbalance}) implies that, if the external parameter is fixed at $\lambda_k$ and is not changed in time, the steady distribution of system $X$ becomes the canonical distribution
\begin{equation}
p_{\rm eq} (x; \lambda_k)= e^{\beta (F(\lambda_k)-E(x; \lambda_k))},
\end{equation}
where $F(\lambda_k) := -\beta^{-1} \ln \sum_x e^{-\beta E(x; \lambda_k)}$ is the free energy.
In fact, it is easy to check that 
\begin{equation}
\sum_{x_k} p (x_{k+1} | x_k; \lambda_k) p_{\rm eq} (x_k; \lambda_k)=p_{\rm eq} (x_{k+1}; \lambda_k).
\end{equation}

It is known that the expression of the detailed balance (\ref{detailedbalance}) is valid for a much broader class of dynamics than the present setup.  In fact, it is known that   Eq.~(\ref{detailedbalance}) is valid for Langevin dynamics even in the presence of nonconservative force~\cite{Seifert2005}.
Moreover, a slightly modified form of  Eq.~(\ref{detailedbalance}) is valid for nonequilibrium dynamics with multiple heat baths at different temperatures~\cite{Jarzynski2}.
Therefore, we regard  Eq.~(\ref{detailedbalance}) as a starting point of the following argument.

We now consider the entropy production, which is the sum of the entropy changes in system $X$ and the heat bath. 
The stochastic entropy change in system $X$ from time $k$ to $k+1$ is given by 
\begin{align}
\Delta s^X_k := s(x_{k+1}) - s(x_k),
\label{stochastic_entropy_X}
\end{align}
where $s (x_k) := - \ln p(x_k)$ is the stochastic Shannon entropy.
The ensemble average of (\ref{stochastic_entropy_X}) gives the change in the Shannon entropy as $\langle \Delta s^X_k \rangle := \langle s(x_{k+1}) \rangle - \langle s(x_k) \rangle$.  The total stochastic entropy change in $X$ from time $1$ to $N$ is given by
\begin{equation}
\Delta s^X := \sum_{k=1}^{N-1 }\Delta s_k^X = s(x_N) - s(x_1),
\end{equation}
which is also written as
\begin{equation}
\Delta s^X = \ln \frac{p(x_1)}{p(x_N)}.
\label{system_trajectory}
\end{equation}

The stochastic entropy change in the heat bath is identified with the heat dissipation into the bath~\cite{Seifert2005}:
\begin{equation}
\Delta s^{\rm bath}_k := -\beta Q_k.
\label{bath_entropy}
\end{equation}
From Eq.~(\ref{detailedbalance1}), Eq.~(\ref{bath_entropy}) can also be rewritten as
\begin{align}
\Delta s^{\rm bath}_k = \ln \frac{p(x_{k+1}|x_k; \lambda_k)}{p(x_k|x_{k+1}; \lambda_k)}.
\label{detailedFT}
\end{align}
The total stochastic entropy change in the heat bath  from time $1$ to $N$ is then given by
\begin{equation}
\Delta s^{\rm bath} := \sum_{k=1}^{N-1} \Delta s^{\rm bath}_k  = -\beta Q,
\end{equation}
which can be rewritten as
\begin{equation}
\Delta s^{\rm bath} = \ln \frac{p(x_N|x_{N-1};\lambda_{N-1}) \cdots p(x_3|x_2; \lambda_2)p(x_2|x_1;\lambda_1)}{p(x_1|x_2;\lambda_1)p(x_2|x_3; \lambda_2)\cdots p(x_{N-1}|x_N;\lambda_{N-1})}.
\label{bath_trajectory}
\end{equation}

The total stochastic entropy production of system $X$ and the heat bath from time $k$ to $k+1$ is then defined as
\begin{equation}
\sigma_k := \Delta s^X_k + \Delta s^{\rm bath}_k,
\end{equation}
and that from time $1$ to $N$ is defined as
\begin{equation}
\sigma := \Delta s^X + \Delta s^{\rm bath}.
\label{EntropyProduction0}
\end{equation}
The entropy production $\langle \sigma \rangle$ is defined as the average of $\sigma$, where $\langle \cdots \rangle$ denotes the ensemble average over probability distribution $p(\mathcal{X})$.
From Eqs.~(\ref{system_trajectory}) and (\ref{bath_trajectory}), we obtain
\begin{equation}
\sigma = \ln \frac{p(x_N|x_{N-1};\lambda_{N-1}) \cdots p(x_3|x_2; \lambda_2)p(x_2|x_1;\lambda_1)p(x_1)}{p(x_1|x_2;\lambda_1)p(x_2|x_3; \lambda_2)\cdots p(x_{N-1}|x_N;\lambda_{N-1}) p(x_N)},
\label{total_trajectory}
\end{equation}
which is sometimes referred to as the detailed fluctuation theorem~\cite{Jarzynski2}.

We discuss the meaning of the probability distributions in the right-hand side of Eq.~(\ref{total_trajectory}).
First,  we recall that the probability distribution of $\mathcal{X}$ is given by
\begin{equation}
p(\mathcal{X}) := p(x_N|x_{N-1};\lambda_{N-1}) \cdots p(x_3|x_2; \lambda_2)p(x_2|x_1;\lambda_1)p(x_1),
\label{notation_F}
\end{equation}
which describes the probability of trajectory $x_1 \to x_2 \to \cdots \to x_N$ with the time evolution of the external parameter $\lambda_1 \to \lambda_2 \to \cdots \lambda_N$.
On the other hand, 
\begin{equation}
 p_{\rm B}(\mathcal{X}) := p(x_1|x_2;\lambda_1)p(x_2|x_3; \lambda_2)\cdots p(x_{N-1}|x_N;\lambda_{N-1}) p(x_N)
 \label{notation_B}
\end{equation}
 is regarded as the probability of the ``backward'' trajectory $x_N \to x_{N-1} \to \cdots \to x_1$ starting from the initial distribution $p(x_N)$, where the time evolution of the external prarameter  is also time-reversed as  $\lambda_N \to \lambda_{N-1} \to \cdots \to \lambda_1$.  In other words, $ p_{\rm B}(\mathcal{X})$ describes the probability of the time-reversal of the original dynamics.
To emphasize this, we introduced suffix ``B'' in  $p_{\rm B}(\mathcal{X})$ that represents ``backward.''  We also write
\begin{equation}
 p_{\rm B}(x_{k-1}|x_k) := p(x_{k-1}|x_k;\lambda_{k-1}).
 \label{notation_b}
\end{equation}
We again stress that $ p_{\rm B}(\mathcal{X})$ is different from the original probability $p(\mathcal{X}) $, but describes the  probability of the time-reversed trajectory with the time-reversed time evolution of the external parameter.
By using notations~(\ref{notation_F}) and (\ref{notation_B}), Eq.~(\ref{total_trajectory}) can be written in a simplified way:
\begin{equation}
\sigma = \ln \frac{p(\mathcal{X})}{p_{\rm B} (\mathcal {X})}.
\label{detailed_FT}
\end{equation}
In Eqs.~(\ref{total_trajectory}) and (\ref{detailed_FT}), the entropy production is determined by the ratio of the probabilities of a trajectory and its time-reversal.  This implies that the entropy production is a measure of irreversibility.  

We consider the second law of thermodynamics, which states that the average entropy production is nonnegative:
\begin{equation}
\langle \sigma \rangle \geq 0.
\label{second_law}
\end{equation}
This is a straightforward consequence of the definition of $\sigma$ as shown below.  We first note that Eq.~(\ref{detailed_FT}) can be rewritten by using the stochastic relative entropy defined in Eq.~(\ref{stochastic_relative}):
\begin{equation}
\sigma = d_{\rm KL} (p( \mathcal{X}) \|  p_B(  \mathcal{X})).
\end{equation}
By taking the ensemble average of $d_{\rm KL} (p( \mathcal{X}) \|  p_B(  \mathcal{X}))$ by the probability distribution $p(\mathcal{X})$, we find that $\langle \sigma \rangle$ is equal to the relative entropy between $p( \mathcal{X})$ and $p_B( \mathcal{X})$:
\begin{equation}
\langle \sigma \rangle = \langle d_{\rm KL} (p( \mathcal{X}) \|  p_B( \mathcal{X})) \rangle =: D(p\| p_{\rm B}),
\end{equation}
which is nonnegative and  implies inequality~(\ref{second_law}).

The second law (\ref{second_law}) can be shown in another way as follows.
We first show that 
\begin{align}
\langle \exp (- \sigma ) \rangle =1,
\label{IFT}
\end{align}
because
\begin{equation}
\langle \exp (- \sigma ) \rangle = \left\langle \frac{p_{\rm B} (\mathcal {X})}{p(\mathcal{X})}  \right\rangle = \sum_{\mathcal X }p(\mathcal{X}) \frac{p_{\rm B} (\mathcal {X})}{p(\mathcal{X})} = \sum_{\mathcal X }p_B(\mathcal{X}) =1.
\end{equation}
Equality~(\ref{IFT}) is called the integral fluctuation theorem~\cite{Jarzynski1,Seifert2005}.   By applying the Jensen inequality, we obtain
\begin{align}
\langle \exp (- \sigma ) \rangle \geq \exp (- \langle  \sigma  \rangle ),
\end{align}
which, along with Eq.~(\ref{IFT}), leads to the second law  (\ref{second_law}).  We note that Eq.~(\ref{IFT}) can be regarded as a special case of Eq.~(\ref{relative_FT}), and the above proof of inequality~(\ref{second_law}) is parallel to the argument below  Eq.~(\ref{relative_FT}).

We next consider the physical meaning of the entropy production for a special case, and relate the entropy production to the work and the free energy.
Suppose that the initial and the final probability distributions are given by the canonical distributions such that $p(x_1) = p_{\rm eq}(x_1; \lambda_1)$ and $p(x_N) = p_{\rm eq} (x_N; \lambda_N)$.  In this case, the stochastic Shannon entropy change is given by
\begin{equation}
\Delta s^X = \ln \frac{p_{\rm eq}(x_1; \lambda_1)}{p_{\rm eq} (x_N; \lambda_N)} = - \beta (\Delta F - \Delta E),
\end{equation}
where $\Delta F := F(\lambda_N) - F(\lambda_1)$ is the free energy change and $\Delta E := E(x_N; \lambda_N) - E(x_1; \lambda_1)$ is the energy change.
Therefore, the stochastic entropy production is given by 
\begin{equation}
\sigma = \Delta s^X - \beta Q = \beta (- \Delta F + \Delta E - Q).
\end{equation}
By using the first law of thermodynamics (\ref{first_law}), we obtain
\begin{equation}
\sigma = \beta (W - \Delta F).
\label{entropy_WF}
\end{equation}
Equality~(\ref{entropy_WF}) gives the energetic interpretation of the entropy production for transitions between equilibrium states.
In this case, the integral fluctuation theorem~(\ref{IFT}) reduces to
\begin{equation}
\langle e^{\beta (\Delta F - W)} \rangle = 1,
\label{Jarzynski}
\end{equation}
which is called the Jarzynski equality~\cite{Jarzynski1}.  It can also be shown that Eq.~(\ref{Jarzynski}) is still valid even when the final distribution is out of equilibrium~\cite{Jarzynski1}.
The second law of thermodynamics (\ref{second_law}) then reduces to
\begin{equation}
\langle W \rangle \geq \Delta F,
\end{equation}
which is a well-known energetic expression of the second law; the free energy increase cannot be larger than the performed work.

\section{Bayesian networks\label{Sec_Bayesian}}

In this section, we review the basic concepts of Bayesian networks~\cite{Minsky, Pearl, Bishop, PearlBook, PearlBook2, JensenBook}, which represent causal structures of stochastic dynamics with directed acyclic graphs.

\begin{figure}[htbp]
 \begin{center}
  \includegraphics[width=50mm]{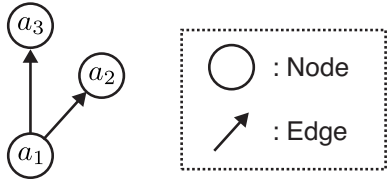}
 \end{center}
 \caption{Example of a simple directed acyclic graph $\mathcal{G} = \{ \mathcal{V}, \mathcal{E} \}$ with $\mathcal{V} = \{ a_1, a_2, a_3 \}$ and $\mathcal{E} = \{ a_1 \to a_2, a_1 \to a_3 \}$.} 
 \label{DAGExampleIto}
\end{figure}

\begin{figure}[htbp]
 \begin{center}
  \includegraphics[width=60mm]{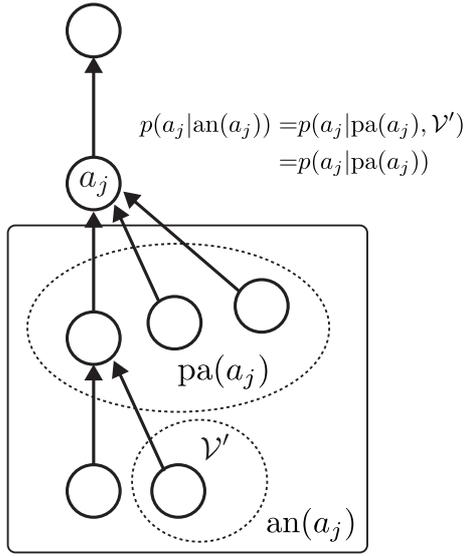}
 \end{center}
 \caption{Schematic of the parents of $a_j$. The set of the parents, ${\rm pa} (a_j)$, is defined as the set of  the nodes that have directed edges toward $a_j$.  This figure also illustrates  the setup of  Eq.~(\ref{localMP}).} 
 \label{LocalMPIto}
\end{figure}

We first define the directed acyclic graph (see also Fig.~\ref{DAGExampleIto}). The directed graph $\mathcal{G} := \{ \mathcal{V}, \mathcal{E} \}$ is given by a finite set of nodes $\mathcal{V}$ and a finite set of directed edges $\mathcal{E}$. We write the set of nodes as 
\begin{equation}
\mathcal{V}= \{ a_1, \dots, a_{N_{\mathcal{V}}}\},
\end{equation}
where $a_j$ is a node and $N_{\mathcal{V}}$ is the number of nodes. 
The set of directed edges $\mathcal{E}$ is given by a subset of all ordered pairs of nodes in $\mathcal{V}$:
\begin{equation}
\mathcal{E} := \{ a_j \to a_{j'} | a_j , a_{j'} \in \mathcal{V}, a_j \neq a_{j'} \}.
\end{equation}
Intuitively, $\mathcal V$ is the set of events, and their causal relationship is represented by  $\mathcal{E}$.
If $(a_j \to a_{j'}) \in \mathcal{E}$, we say that $a_j$ is a {\it parent} of $a_{j'}$ (or equivalently, $a_{j'}$ is a {\it child} of $a_{j}$). We write as ${\rm pa} (a_j)$ the set of parents of $a_{j}$ (see also Fig. \ref{LocalMPIto}): 
\begin{equation}
{\rm pa} (a_j) := \{ a_{j'}| ( a_{j'} \to a_j ) \in \mathcal{E} \}.
\end{equation}

A directed graph is called {\it acyclic} if   $\mathcal{E}$ does not include any directed cyclic path.  In other words, a directed graph is {\it cyclic} if there exists $(j, j^{(1)}, j^{(2)}, \cdots, j^{(n)})$ such that  $\{ a_j \to a_{j^{(1)}}, a_{j^{(1)}} \to a_{j^{(2)}}, \dots, a_{j^{(n-1)}} \to a_{j^{(n)}} , a_{j^{(n)}} \to a_{j} \} \subset \mathcal{E}$; otherwise, it is acyclic.
The acyclic property implies that the causal structure does not include any ``time loop.''
If a directed graph is acyclic, we can define the concept of {\it topological ordering}.
An ordering of $\mathcal{V}$, written as $( a_1, a_2, \dots, a_{N_{\mathcal{V}}})$, is called topological ordering, if 
 $a_j$ is not a parent of $a_{j'}$ for $j>j'$. 
We then define the set of {\it ancestors} of $a_j$ by ${\rm an} (a_j) := \{ a_{j-1}, \dots, a_1 \}$ (${\rm an} (a_1) := \emptyset$). 
We note that a topological ordering is not necessary unique.

We show a simple example of a directed acyclic graph $\mathcal{G} = \{ \mathcal{V}, \mathcal{E} \}$ with $\mathcal{V} = \{ a_1, a_2, a_3 \}$ and $\mathcal{E} = \{ a_1 \to a_2, a_1 \to a_3 \}$ in Fig. \ref{DAGExampleIto}. 
A node is described by a circle with variable $a_j$, and a directed edge is described by a directed arrow between two nodes. In Fig. \ref{DAGExampleIto}, the sets of parents are given by ${\rm pa} (a_1) = \emptyset$, ${\rm pa} (a_2)= \{ a_1 \}$ and ${\rm pa} (a_3)= \{ a_1 \}$, where $\emptyset$ denotes the empty set. In this case, we have two topological orderings: $\{ a_1, a_2, a_3 \}$ and $\{ a_1, a_3, a_2 \}$.

We next consider a probability distribution on a directed acyclic graph  $\mathcal{G} = \{ \mathcal{V}, \mathcal{E}\}$, which is a key concept for  Bayesian networks.
A directed edge $a_j \to a_{j'} \in \mathcal{E}$ on a Bayesian network represents the probabilistic dependence (i.e., causal relationship) between two nodes $a_j$ and $a_{j'}$. 
Therefore,  variable $a_j$ only depends on its parents ${\rm pa} (a_j)$.
The causal relationship can be described by the conditional probability of $a_j$ under the condition of  ${\rm pa} (a_j)$, written as $p( a_j | {\rm pa} (a_j))$.
If  ${\rm pa} (a_j) = \emptyset$, $p(a_j) := p( a_j | \emptyset)$ is just the probability of $a_j$.
The joint probability  distribution of all the nodes in a Bayesian network is then defined as
\begin{eqnarray}
p(\mathcal{V}) := \prod^{\mathcal{N}_{V}}_{j=1}  p( a_j | {\rm pa} (a_j)),
\label{ChainBayes}
\end{eqnarray}
which implies that the probability of a node is only determined by its parents.  This definition represents the causal structure of Bayesian networks; the cause of a node is given by its parents.

\begin{figure}[htbp]
 \begin{center}
  \includegraphics[width=70mm]{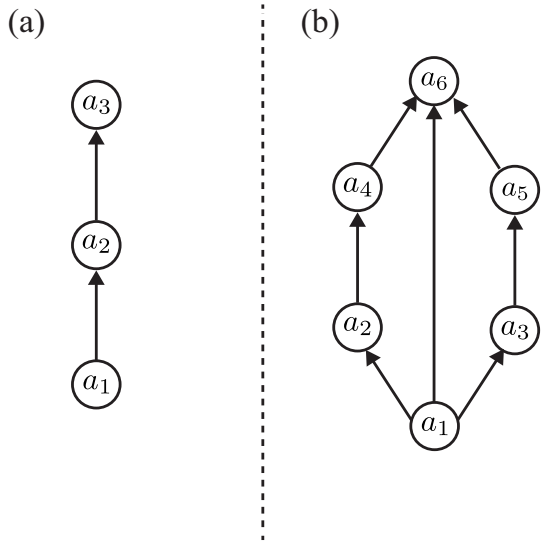}
 \end{center}
 \caption{Simple examples of Bayesian networks.} 
 \label{examples}
\end{figure}

In Fig. \ref{examples}, we show two simple examples of Bayesian networks.  For Fig. \ref{examples} (a), the joint distribution is given by
\begin{equation}
p(a_1, a_2, a_3) := p(a_3 | a_2) p(a_2 | a_1) p(a_1),
\end{equation}
which describes a simple Markovian process.  Figure \ref{examples} (b) is a little less trivial,  whose joint distribution is given by
\begin{equation}
p(a_1, a_2, \cdots, a_6) = p(a_6 | a_1,a_4,a_5) p(a_5 | a_3) p(a_4 | a_2) p(a_3 | a_1) p(a_2 | a_1) p(a_1).
\end{equation}

For any subset of nodes $\mathcal{A} \subseteq \mathcal{V}$, the probability distribution on $\mathcal A$ is given by 
\begin{equation}
p(\mathcal A) = \sum_{\mathcal V \setminus \mathcal A } p(\mathcal V).
\end{equation} 
For $\mathcal{A}, \mathcal{A}' \subseteq \mathcal{V}$, the joint probability distribution is given by
\begin{equation}
p(\mathcal A, \mathcal A') = \sum_{\mathcal V \setminus ( \mathcal A \cup \mathcal A' )} p(\mathcal V).
\end{equation} 
The conditional probability is then given by the Bayes rule:
\begin{equation}
p(\mathcal A | \mathcal A') = \frac{p(\mathcal A, \mathcal A')}{p(\mathcal A')}.
\label{Bayes_rule}
\end{equation}
Let $A (\mathcal V)$ be a probability variable that depends on nodes in $\mathcal V$.  The ensemble average of $A (\mathcal V)$ is defined as
\begin{equation}
\langle A (\mathcal V) \rangle := \sum_{\mathcal V} A (\mathcal V) p (\mathcal V).
\label{ensemble_average}
\end{equation}
In particular, if $A$ depends only on $\mathcal A \subseteq \mathcal V$, Eq.~(\ref{ensemble_average}) reduces to
\begin{equation}
\langle A (\mathcal A) \rangle := \sum_{\mathcal A} A (\mathcal A) p (\mathcal A) = \sum_{\mathcal V} A (\mathcal A) p (\mathcal V).
\end{equation}

We note that $p( a_j | {\rm an} (a_j)) = p( a_j | {\rm pa} (a_j))$ holds by definition, which implies that any probability variable directly depends on the nearest ancestors (i.e., parents).  This is consistent with the description of directed acyclic graphs.   In  general,
we have
\begin{equation}
p(a_j| {\rm pa} (a_j), \mathcal{V}' ) = p(a_j|{\rm pa}(a_j))
\label{localMP}
\end{equation}
for any $\mathcal{V}' \subseteq \{ {\rm an}(a_j) \setminus {\rm pa} (a_j) \}$ (see also Fig.~\ref{LocalMPIto}).

\section{Information thermodynamics on Bayesian networks\label{Sec_Information1}}

We now discuss a general framework of stochastic thermodynamic for complex dynamics described by Bayesian networks~\cite{Ito}, where system $X$ is in contact with systems $C$ in addition to the heat bath. 
 In particular, we derive the generalized second law of thermodynamics, which states that  the entropy production  is bounded by an informational quantity that consists of the initial and final mutual between $X$ and $C$, and the transfer entropy from $X$ to $C$.

 \subsection{Setup\label{Sec_Setup2}}
 
 First of all, we discuss how Bayesian networks represent the causal relationships in physical dynamics. We consider a situation that several physical systems interact with each other and stochastically evolve in time.
 A probability variable associated with a node, $a_{j} \in\mathcal{V}$, represents a state of one of the systems at a particular time.  We assume that the topological ordering $(a_1, \dots, a_{N_{\mathcal{V}}})$ describes the time ordering;  the time of state $a_{j}$ should not be later than the time of state $a_{j+1}$. 
This assumption does not exclude a situation that $a_{j}$ and $a_{j+1}$ can be states of different systems at the same time.
 Each edge in $\mathcal{E}$ describes the causal relationship between states of the systems at different times.
Correspondingly, the conditional probability $p( a_j | {\rm pa} (a_j))$ characterizes the stochastic dynamics.
The joint probability $p(\mathcal{V})$ represents the probability of trajectories of the whole system.

\begin{figure}[htbp]
 \begin{center}
  \includegraphics[width=50mm]{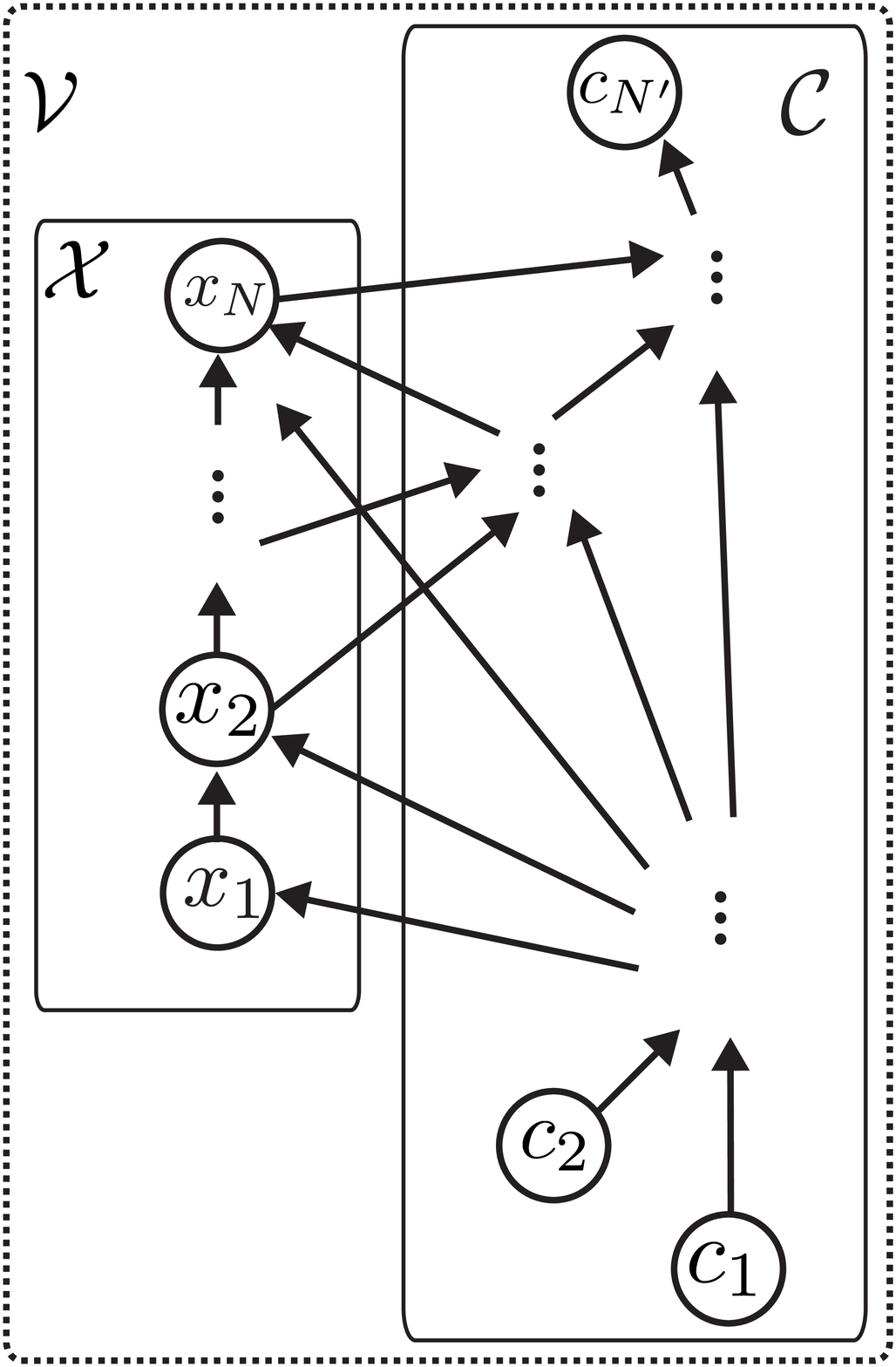}
 \end{center}
 \caption{Schematic of the physical setup of Bayesian networks. The time evolution of system $X$ is given by the sequence of nodes $\mathcal{X} = \{ x_1, \cdots, x_N \}$, and the time evolution of $C$ is given by $\mathcal {C} := \mathcal{V} \setminus \mathcal{X}= \{ c_1, \cdots, c_{N'} \}$. } 
 \label{CIto}
\end{figure}

 We focus on a particular system $X$, whose time evolution is described by a set of nodes.
Let $\mathcal{X} := \{ x_1, \dots, x_N  \} \subseteq \mathcal{V}$ be the set of nodes that describe states of $X$, and let $(x_1, \dots, x_N )$ be the topological ordering of the elements of $\mathcal{X}$, where we refer to the suffixes as ``time.''
A probability variable $x_k$ in $\mathcal{X}$ describes the state of system $X$ at time $k$. 
We assume that there is a causal relationship between $x_k$ and $x_{k+1}$ such that 
 \begin{eqnarray}
x_k \in  {\rm pa} (x_{k+1}).
\label{markovX1}
\end{eqnarray}
For simplicity, we also assume that
 \begin{eqnarray}
{\rm pa} (x_{k+1}) \cap \mathcal{X} = \{ x_k \},
\label{markovX2}
\end{eqnarray}
which does not exclude the situation that there are nodes in ${\rm pa} (x_{k+1})$ outside of $\mathcal X$ (see  Fig.~\ref{CIto}).

We next consider the systems other than $X$, which we refer to as $C$.  The states of $C$ are given by  the nodes  in set $\mathcal{C} := \mathcal{V} \setminus \mathcal{X}$  (see also Fig.~\ref{CIto}).
Let $(c_1,c_2, \cdots, c_{N'})$ be the topological ordering of $\mathcal C$, where we again refer to the suffixes as ``time.''
A probability variable $c_l$ describes the state of $C$ at time $l$.
Since $\mathcal{V} = \mathcal {X} \cup \mathcal {E}$, we can define an joint topological ordering of $\mathcal{V}$ as
\begin{equation}
( c_1,\dots, c_{l^{(1)}}, x_1, c_{l^{(1)}+1}, \dots, c_{l^{(2)}}, x_2, c_{l^{(2)}+1}, \dots, \dots c_{l^{(N)}}, x_N, c_{l^{(N)}+1} ,\dots, c_{N'} ),
\end{equation}
where the ordering $(c_1,\dots, c_{l^{(1)}}, \dots, c_{l^{(2)}}, \dots, c_{N'})$ is the same as the ordering $(c_1, c_2, \dots, c_{N'})$.
The joint probability distribution $p(\mathcal{X}, \mathcal{C})$ can be obtained from Eq.~(\ref{ChainBayes}):
\begin{eqnarray}
p(\mathcal {X}, \mathcal{C}) = \prod^{N}_{k=1}  p( x_k | {\rm pa} (x_k))  \prod^{N'}_{l=1}  p( c_l | {\rm pa} (c_l)),
\label{ChainBayes2}
\end{eqnarray}
where the conditional probability $ p( x_{k+1} | {\rm pa} (x_{k+1}))$ represents the transition probability of  system $X$ from time $k$ to $k+1$. 
We note that the dynamics in $\mathcal{V}$ can be non-Markovian due to the non-Markovian property of  $C$.
We summarize the notations in Table~\ref{Table_notations}.

\begin{table}[htb]
\begin{center}
\begin{tabular}{|l|l|} \hline
Notation & Meaning  \\ \hline \hline
${\rm pa} (a)$ (Parents of $a$) &   Set of nodes that have causal relationship to node $a$.  \\ \hline
${\rm an}(a)$ (Ancestors of $a$) & Set of nodes before node $a$ in the topological ordering.\\ \hline \hline
$x_k$ & State of system $X$ at time $k$. \\ \hline
$\mathcal{X} := \{ x_1, \cdots, x_N \}$ & Set of states of system $X$.  \\ \hline
$\mathcal{C} := \{ c_1, \cdots, c_{N'} \}$ & Set of states of other systems $C$. \\ \hline
$\mathcal{C}' := {\rm an}(x_N) \cap \mathcal{C}$ &Set of the ancestors of $x_N$ in $C$. \\ \hline
${\rm pa}_{\mathcal X} (c_l) := {\rm pa} (c_l) \cap \mathcal{X}$ & Set of the parents of $c_l$ in $\mathcal{X}$. \\ \hline
${\rm pa}_{\mathcal C} (c_l) := {\rm pa} (c_l) \cap \mathcal{C}$ & Set of the parents of $c_l$ in $\mathcal{C}$. \\ \hline
$\mathcal{B}_{k+1} := {\rm pa} (x_{k+1}) \setminus \{ x_k \}$ & Set of parents of $x_{k+1}$ outside of $\mathcal{X}$. \\ \hline \hline
$I_{\rm ini} := I(x_1 : {\rm pa} (x_1))$ & Initial mutual information between  $X$ and $C$. \\ \hline
$I_l^{\rm tr} := I(c_l: {\rm pa}_{\mathcal X} (c_l)| c_1, \cdots, c_{l-1})$ & Transfer entropy from  $X$ to $c_l$. \\ \hline
$I^{\rm tr} := \sum_{l \ : \ c_l \in \mathcal{C}'} I_l^{\rm tr}$ & Total transfer entropy from  $X$ to $C$. \\ \hline
$I_{\rm fin} := I(x_N : \mathcal{C}')$ & Final mutual information between  $X$ and $C$. \\ \hline
$\Theta := I_{\rm fin} - I^{\rm tr} - I_{\rm ini}$ & $-\langle \Theta \rangle$ is the available information about $X$ obtained by $C$. \\ \hline
$\sigma$ & Stochastic entropy change in $X$ and the heat bath. \\ \hline 
\end{tabular}
\caption{Summary of notations.}
\label{Table_notations}
\end{center}
\end{table}

\subsection{Information contents on Bayesian networks\label{Sec_Information2}}

We consider information contents on Bayesian networks; the initial and the final mutual information between $X$ and $C$, and the transfer entropy from $X$ to  $C$.

\begin{figure}[htbp]
 \begin{center}
  \includegraphics[width=140mm]{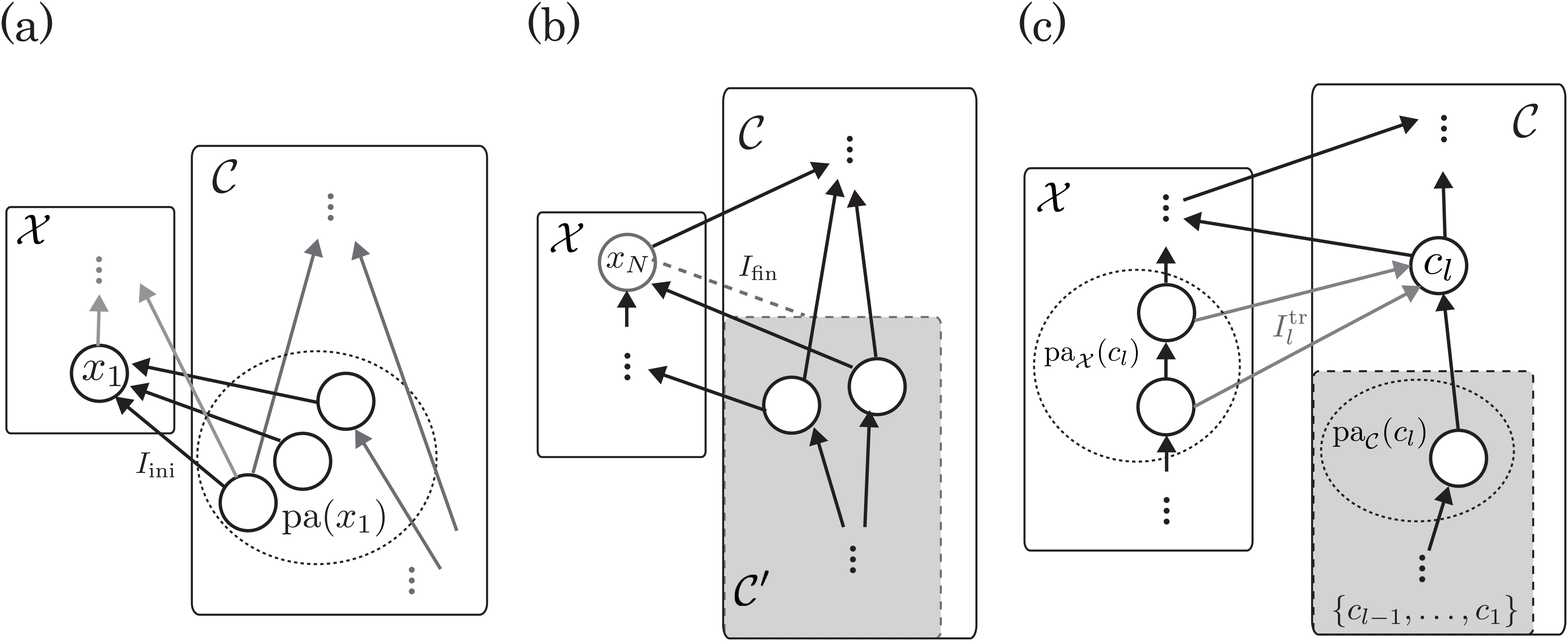}
 \end{center}
 \caption{Schematics of informational quantities on Bayesian networks. (a) The initial correlation between $x_1$ and ${\rm pa} (x_1)$.  (b) The final correlation between $x_N$ and $\mathcal{C}'$.  (c) The transfer entropy from from $X$ to $c_l$.} 
 \label{informationalquantity}
\end{figure}

We first consider the initial correlation of the dynamics.
The initial state $x_1$ of $X$ is initially correlated with its parents ${\rm pa}(x_1) \subseteq \mathcal{C}$.
The initial correlation between  system $X$ and $C$ is then characterized by the mutual information between $x_1$ and ${\rm pa}(x_1)$.  The corresponding stochastic mutual information is given by (see also Fig.~\ref{informationalquantity} (a))
\begin{eqnarray}
I_{\rm ini} := I(x_1 : {\rm pa} (x_1)). 
\label{inicorr}
\end{eqnarray}
Its ensemble average $\langle I_{\rm ini} \rangle$ is the mutual information of the initial correlation.  It vanishes if and only if $p(x_1 |{\rm pa}(x_1)) = p(x_1)$, or equivalently, ${\rm pa}(x_1) = \emptyset$.


We next consider the final correlation of the dynamics. 
The final state of $X$ is given by $x_N \in \mathcal X$, which is correlated with its ancestors ${\rm an}(x_N)$.
The final correlation between system $X$ and $C$ is then characterized by the mutual information between $x_N$ and  $\mathcal{C}' := {\rm an}(x_N) \cap \mathcal{C}$. 
The corresponding stochastic  mutual information is given by (see also Fig.~\ref{informationalquantity} (b))
\begin{eqnarray}
I_{\rm fin} := I(x_N : \mathcal{C}'). 
\label{fincorr}
\end{eqnarray}
Its ensemble average $\langle I_{\rm fin} \rangle$ is the mutual information of the final correlation.  It vanishes  if and only if $p(x_N| \mathcal{C}') = p(x_N)$.


We next consider the transfer entropy from $X$ to $C$ during the dynamics.  The transfer entropy on Bayesian networks has been discussed in Ref.~\cite{Polani}. We here focus on the role of the transfer entropy on Bayesian networks in terms of information thermodynamics.

Let $c_l \in \mathcal C$.
Let  ${\rm pa}_{\mathcal X} (c_l) := {\rm pa}(c_l) \cap \mathcal{X}$  be the set of the parents of $c_l$ in $\mathcal{X}$, and ${\rm pa}_{\mathcal{C}} (c_l) := {\rm pa}(c_l) \cap \mathcal{C}$ be the set of the parents of $c_l$ in $\mathcal{C}$  (see also Fig.~\ref{informationalquantity} (c)). We note that ${\rm pa}_{\mathcal X} (c_l)  \cup {\rm pa}_{\mathcal{C}} (c_l) = {\rm pa} (c_l) $ and  ${\rm pa}_{\mathcal X} (c_l)  \cap {\rm pa}_{\mathcal{C}} (c_l) = \emptyset$.
We then have
\begin{equation}
\begin{split}
p(c_l |{\rm pa}(c_l) ) &= p(c_l |{\rm pa}_{\mathcal X} (c_l), {\rm pa}_{\mathcal{C}} (c_l)) \\
&= p(c_l |{\rm pa}_{\mathcal X} (c_l), c_1, \dots, c_{l-1}),
\end{split}
\end{equation}
where we used Eq.~(\ref{localMP}) with $\mathcal{V}' = \{c_1, \cdots, c_{l-1}  \} \setminus  {\rm pa}_{\mathcal{C}} (c_l)$.


The transfer entropy from system $X$ to state $c_l$ is defined as the conditional mutual information between $c_l$ and  ${\rm pa}_{\mathcal X} (c_l) $ under the condition of  $\{ c_1 , \dots, c_{l-1} \}$. 
The  corresponding stochastic transfer entropy is given by
\begin{equation}
\begin{split}
I_l^{\rm tr}  &:=  I (c_l : {\rm pa}_{\mathcal X} (c_l) |c_1, \dots, c_{l-1} )\\
&:= \ln \frac{p(c_l, {\rm pa}_{\mathcal X} (c_l) |  c_1, \cdots, c_{l-1})}{p(c_l | c_1, \cdots, c_{l-1}) p( {\rm pa}_{\mathcal X} (c_l) | c_1, \cdots, c_{l-1})}\\
&= \ln \frac{\ln p(c_l|{\rm pa}_{\mathcal X} (c_l), c_1, \cdots, c_{l-1} )}{p(c_l|c_1, \cdots, c_{l-1})}.
\end{split}
\label{transfer}
\end{equation}
It can also be rewritten by using the conditional stochastic Shannon entropy:
\begin{equation}
I_l^{\rm tr} = s(c_l|c_1, \cdots, c_{l-1} ) - s(c_l|{\rm pa}_{\mathcal X} (c_l), c_1, \cdots, c_{l-1}),
\end{equation}
which is analogous to Eq.~(\ref{transfer_conditoinal_Shannon}).
The  ensemble average of $I_l^{\rm tr}$ is the transfer entropy $\langle I_l^{\rm tr} \rangle$, which describes the amount of information about $X$ that is newly obtained by $C$ at time $l$.   $\langle I_l^{\rm tr} \rangle$ is nonnegative from the definition, and is zero if and only if $\ln p(c_l|{\rm pa}_{\mathcal X} (c_l), c_1, \cdots, c_{l-1} )  =\ln p(c_l|c_1, \cdots, c_{l-1} )$, or equivalently ${\rm pa}_{\mathcal X} (c_l) = \emptyset$.
The total transfer entropy from $X$ to $C$ during the dynamics from $x_1$ to $x_N$ is then given by 
\begin{equation}
I^{\rm tr} := \sum_{l \ : \ c_l \in \mathcal{C}'}  I_l^{\rm tr}.
\end{equation}

By summing up the foregoing information contents, we introduce a key informational quantity $\Theta$:
\begin{eqnarray}
\Theta &:=& I_{\rm fin} -I^{\rm tr} - I_{\rm ini},
\label{netinfoflow}
\end{eqnarray}
which plays a crucial role in the generalized second law that will be discussed in Sec.~\ref{Sec_Information1}.
Here, the minus of the ensemble average of $\Theta$ (i.e., $- \langle \Theta \rangle$) characterizes the available information about $X$ obtained by $C$ during the dynamics from $x_1$ to $x_N$ (see also Fig. \ref{informationalquantity}). 

\subsection{Entropy production\label{Sec_Entropy2}}

We next define the entropy production that is defined as the sum of the entropy changes in system $X$ and the heat bath.  While the key idea of the definition is the same as the case for the Markovian dynamics discussed in the previous section, a careful argument is necessary for the entropy production on Bayesian networks, because of the presence of other systems $C$.

\begin{figure}[htbp]
 \begin{center}
  \includegraphics[width=80mm]{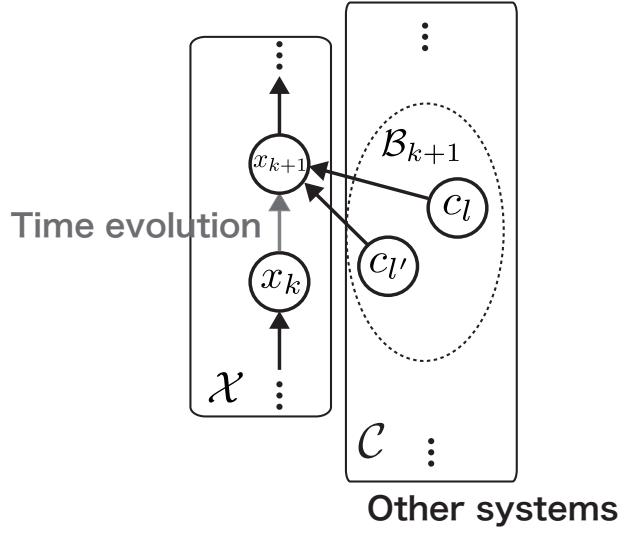}
 \end{center}
 \caption{Schematic of $\mathcal{B}_{k+1}$. } 
 \label{BIto}
\end{figure}

We consider the subset of probability variables in $C$ (i.e., nodes in $\mathcal C$) that affect the time evolution of $X$ from time $k$ to $k+1$, which is defined as (see Fig. \ref{BIto})
\begin{eqnarray}
\mathcal{B}_{k+1} := {\rm pa} (x_{k+1}) \setminus \{ x_k \} \subseteq \mathcal{C}.
\end{eqnarray}
The transition probability of $X$ from time $k$ to $k+1$ is then written as 
\begin{eqnarray}
p(x_{k+1}|{\rm pa}(x_{k+1}) ) = p(x_{k+1}|x_k, \mathcal{B}_{k+1}).
\label{transprob}
\end{eqnarray}

We note that $p(x_{k+1}|x_k, \mathcal{B}_{k+1})$ describes the transition probability from $x_k$ to $x_{k+1}$ under the condition that the states of $C$ that affect $X$ are given by $\mathcal{B}_{k+1}$.
We define the functional form of $p(x_{k+1}|x_k, \mathcal{B}_{k+1})$ with arguments $(x_{k+1}, x_k, \mathcal{B}_{k+1})$ by
\begin{equation}
f(x_{k+1}, x_k, \mathcal{B}_{k+1}) := p(x_{k+1}|x_k, \mathcal{B}_{k+1}).
\end{equation}
We then define the  backward transition probability as
\begin{equation}
p_{\rm B}(x_{k}|x_{k+1}, \mathcal{B}_{k+1}) := f(x_{k}, x_{k+1}, \mathcal{B}_{k+1}),
\label{backward_general}
\end{equation}
which describes the transition probability from $x_{k+1}$ to $x_k$ under the same condition $\mathcal{B}_{k+1}$ as the forward process.

Here, $p_{\rm B}(x_{k}|x_{k+1}, \mathcal{B}_{k+1})$ is different from the conditional probability $p(x_{k}|x_{k+1}, \mathcal{B}_{k+1}) := p(x_{k}, x_{k+1}, \mathcal{B}_{k+1})/ p(x_{k+1}, \mathcal{B}_{k+1})$, which is obtained from the Bayes rule (\ref{Bayes_rule}) of the Bayesian network.
To emphasize the difference, we used the suffix ``B'' that represents ``backward.'' 
We note that $p_{\rm B}(x_{k}|x_{k+1}, \mathcal{B}_{k+1})$ is analogous to $p(x_k | x_{k+1}; \lambda_k)$ in Eq.~(\ref{notation_b}) of Sec.~\ref{Sec_Stochastic}, by replacing $\lambda_k$ by $\mathcal{B}_{k+1}$.
In fact, in many situations, we can assume that external parameter $\lambda_k$ is determined by $\mathcal{B}_{k+1}$; a typical case is feedback control as will be discussed in Sec.~\ref{Sec_E_Feedback} and \ref{Sec_E_Repeated}.
We also note that the backward probability $p_{\rm B}(x_{k}|x_{k+1}, \mathcal{B}_{k+1})$ can be defined even in the presence of odd variables like momentum, by slightly modifying definition~(\ref{backward_general}).

We then define the entropy change in the heat bath from time $k$ to $k+1$ in the form of Eq.~(\ref{detailedFT}):
\begin{eqnarray}
\Delta s_k^{\rm bath} := \ln \frac{p(x_{k+1}|x_k, \mathcal{B}_{k+1})}{p_{\rm B} (x_k | x_{k+1} , \mathcal{B}_{k+1})}.
\label{detailedFT2}
\end{eqnarray}
We note that $\Delta s_k^{\rm bath}$ can be identified with $-\beta Q_k$ in many situations.  In fact, as mentioned above, if $\mathcal{B}_{k+1}$ affects $x_k$ only through the external parameter, Eq.~(\ref{detailedFT2}) is equivalent to (\ref{detailedFT}).  In such a case, we can show that $\Delta s_k^{\rm bath}= -\beta Q_k$ as discussed in Sec.~\ref{Sec_Stochastic}.
The entropy change in the heat bath from time $1$ to $N$ is then given by
\begin{equation}
\begin{split}
\Delta s^{\rm bath} &:= \sum_{k=1}^{N-1} \Delta s_k^{\rm bath} \\
&= \ln \frac{p(x_N|x_{N-1}, \mathcal{B}_N) \cdots p(x_3|x_2, \mathcal{B}_3)p(x_2|x_1, \mathcal{B}_2)}{p_{\rm B}(x_1|x_2, \mathcal{B}_2)p_{\rm B}(x_2|x_3, \mathcal{B}_3)\cdots p_{\rm B}(x_{N-1}|x_N, \mathcal{B}_N)},
\end{split}
\end{equation}
which is analogous to Eq.~(\ref{bath_trajectory}).
The total entropy change in $X$ and the heat bath from time $1$ to $N$ is then defined as
\begin{align}
\sigma :=  \Delta s^{X}  +\Delta s^{\rm bath}, 
\label{EntropyProduction}
\end{align}
which is also written as
\begin{equation}
\sigma = \ln \frac{p(x_N|x_{N-1}, \mathcal{B}_N) \cdots p(x_3|x_2, \mathcal{B}_3)p(x_2|x_1, \mathcal{B}_2)p(x_1)}{p_{\rm B}(x_1|x_2, \mathcal{B}_2)p_{\rm B}(x_2|x_3, \mathcal{B}_3)\cdots p_{\rm B}(x_{N-1}|x_N, \mathcal{B}_N) p(x_N)}.
\label{Bayesian_sigma}
\end{equation}

\subsection{Generalized second law\label{Sec_Generalized}}

We now consider the relationship between the second law of thermodynamics and informational quantities. The lower bound of the entropy change in system $X$ and the heat bath is given by $\langle \Theta \rangle$:
\begin{equation}
\langle \sigma \rangle \geq \langle \Theta \rangle,
\label{GSL}
\end{equation}
or equivalently,
\begin{equation}
\langle \sigma \rangle \geq  \langle I_{\rm fin} \rangle-  \langle  I^{\rm tr} \rangle -\langle I_{\rm ini} \rangle,
\end{equation}
which is the generalized second law of thermodynamics on Bayesian networks.

The proof of the generalized second law (\ref{GSL}) is as follows.
We first show that  $\sigma -\Theta $ can be rewritten as the stochastic relative entropy:
\begin{equation}
\begin{split}
\sigma  - \Theta =& \ln \left[ \frac{p(x_1) }{p(x_N)} \prod_{k=1}^{N-1}   \frac{p(x_{k+1}|x_k, \mathcal{B}_{k+1})}{p_B (x_k | x_{k+1} , \mathcal{B}_{k+1})} \right]  - \ln \frac{p(x_N, \mathcal{C}' )}{p(x_N)p( \mathcal{C}' )} \\
& +  \ln \frac{p(x_1 |{\rm pa} (x_1))}{ p(x_1)}+  \ln \left[ \prod_{l \ : \ c_l \in \mathcal{C}'} \frac{p(c_l|{\rm pa} (c_l))}{ p(c_l|c_1, \dots, c_{l-1} )} \right]  \\
=& \ln \frac{\prod_{k=1}^{N} p(x_{k}|{\rm pa} (x_k)) \prod_{l \ : \ c_l \in \mathcal{C}' } p(c_l|{\rm pa} (c_l))  }{\prod_{k=1}^{N-1} p_B (x_k | x_{k+1} , \mathcal{B}_{k+1}) p(x_N, \mathcal{C}' ) } \\
=& d_{\rm KL} (p(\mathcal{V})  \| p_{\rm B} (\mathcal{V}) ),
\end{split}
\label{relativeE}
\end{equation}
where we  defined
\begin{eqnarray}
p_{\rm B} (\mathcal{V}) :=p(x_N, \mathcal{C}' ) \prod_{k=1}^{N-1} p_B (x_k | x_{k+1} , \mathcal{B}_{k+1})  \prod_{l \ : \ c_l \notin \mathcal{C}', c_l \in \mathcal{C} } p (c_l | {\rm pa} (c_l)).
\end{eqnarray}
We can confirm that $p_{\rm B} (\mathcal{V})$ is normalized, and can be regarded  as a probability distribution:
\begin{equation}
\begin{split}
\sum_{\mathcal{V}} p_{\rm B} (\mathcal{V}) &= \sum_{X, \mathcal{C}'}  p(x_N, \mathcal{C}' ) \prod_{k=1}^{N-1} p_B (x_k | x_{k+1} , \mathcal{B}_{k+1})  \\
=& \sum_{x_N , \mathcal{C}'}  p(x_N, \mathcal{C}' )  \\
=& 1,
\end{split}
\label{normalization_B}
\end{equation}
where we used $x_N \in \mathcal{X}$, $\mathcal{B}_{k+1} \subseteq \mathcal{C}'$, and  $\sum_{x_k} p_B (x_k | x_{k+1} , \mathcal{B}_{k+1})=1$. 
From Eq.~(\ref{relativeE}) and the nonnegativity of the relative entropy, we show that the ensemble average of $\sigma  - \Theta $ is nonnegative:
\begin{equation}
\langle \sigma  - \Theta  \rangle = D_{\rm KL} (p  \|  p_{\rm B} ) \geq 0,
\label{network_relative}
\end{equation}
which implies the generalized second law  (\ref{GSL}). 
The equality in (\ref{GSL}) holds if and only if $p(\mathcal{V}) =  p_{\rm B}(\mathcal{V})$.

We consider the integral fluctuation theorem corresponding to inequality~(\ref{GSL}).
From Eq.~(\ref{relative_FT}) for the stochastic relative entropy, we have
\begin{equation}
\langle e^{-d_{\rm KL} (p(\mathcal{V})  \| p_{\rm B} (\mathcal{V}) )} \rangle = 1,
\end{equation}
or equivalently,
\begin{equation}
\langle e^{- \sigma + \Theta } \rangle =1.
\label{network_IFT}
\end{equation}
This is the generalized integral fluctuation theorem for Bayesian networks.  By applying the Jensen inequality to Eq.~(\ref{network_IFT}), we again obtain inequality~(\ref{GSL}).

We note that, from inequality~(\ref{GSL}) and $ \langle I_{\rm fin} \rangle \geq 0$, we obtain a weaker bound of the entropy production:
\begin{eqnarray}
\langle \sigma \rangle \geq - \langle  I^{\rm tr}  \rangle -\langle I_{\rm ini} \rangle.
\end{eqnarray}
This weaker inequality can also be rewritten as the nonnegativity of the relative entropy $D_{\rm KL} (p  \| \tilde{p}_{\rm B}  ) \geq 0$, where the probability $\tilde{p}_{\rm B} (\mathcal{V})$ is defined as
\begin{eqnarray}
\tilde{p}_{\rm B} (\mathcal{V}) :=p(x_N) p( \mathcal{C}' ) \prod_{k=1}^{N-1} p_B (x_k | x_{k+1} , \mathcal{B}_{k+1})  \prod_{l \ : \ c_l \notin \mathcal{C}', c_l \in \mathcal{C} } p (c_l | {\rm pa} (c_l)).
\end{eqnarray}
The corresponding integral fluctuation theorem is given by
\begin{align}
\langle e^{ - \sigma-   I^{\rm tr}  - I_{\rm ini} } \rangle = 1.
\end{align}

\section{Examples\label{Sec_Examples}}
In the following, we illustrate special examples, and discuss the physical meaning of the generalized second law (\ref{GSL}).

\subsection{Example 1: Markov chain\label{Sec_E_Markov}}

\begin{figure}[htbp]
 \begin{center}
  \includegraphics[width=140mm]{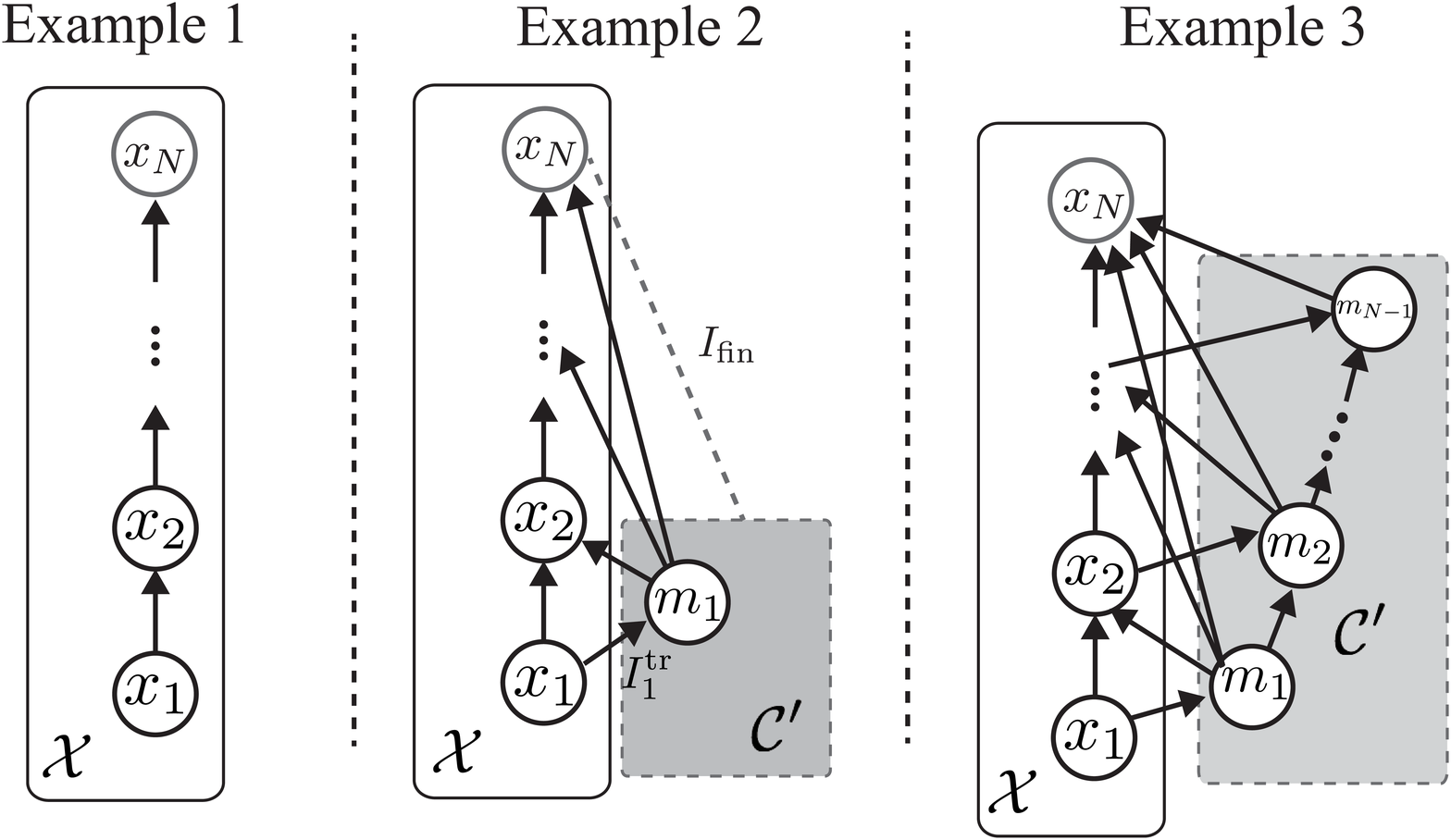}
 \end{center}
 \caption{Bayesian networks of the examples.  Example 1: Markov chain. Example 2: Feedback control with a single measurement. Example 3: Repeated feedback control with multiple adaptive measurements.} 
 \label{example123}
\end{figure}

As the simplest example, we revisit the Markovian dynamics discussed in Sec.~\ref{Sec_Stochastic} from the viewpoint of Bayesian networks.
In this case,  $\mathcal{V} = \mathcal{X} = \{ x_1, \dots, x_N\}$ and $\mathcal{C} = \emptyset$. The Markovian property is characterized by ${\rm pa} (x_k) = \{ x_{k-1}\}$ with $k\geq 2$, and ${\rm pa} (x_1) = \emptyset$ (see also Fig.~\ref{example123}).
Since  $\mathcal{B}_{k+1} := {\rm pa} (x_{k+1}) \setminus \{ x_k\} = \emptyset$,  the entropy production~(\ref{EntropyProduction}) is equivalent to Eq.~(\ref{EntropyProduction0}).
From $\mathcal{C} = \emptyset$, we have $I_{\rm fin} =0$, $I_{\rm ini} = 0$, $ I_l^{\rm tr} =0$, and therefore $\Theta = 0$. 
Therefore, the definition of $\sigma$ in Eq.~(\ref{Bayesian_sigma}) reduces to Eq.~(\ref{detailed_FT}), and the generalized second law (\ref{GSL}) just reduces to $\langle \sigma \rangle \geq 0$.

\subsection{Example 2: Feedback control with a single measurement\label{Sec_E_Feedback}}

We consider the system under feedback control with a single measurement as is the case for the Szilard engine.
In this case, system $X$ is the measured system, and the other system $C$ is a memory that stores the measurement outcome.  

At time $k=1$, a measurement on state $x_1$ is performed, and the obtained outcome is stored in memory state $m_1$.
The probability of outcome $m_1$ under the condition of state $x_1$ is denoted by $p(m_1|x_1)$, which characterizes the measurement error.  If  $p(m_1|x_1)$ is the delta function  $\delta_{m_1, x_1}$, the measurement is error-free.
After the measurement, the time evolution of  $X$ is affected by $m_1$ such that the transition probability of $X$ from time $k$ to $k+1$ is given by $p(x_{k+1}|x_k, m_1)$ ($k=1, 2, \cdots, N-1$), which is feedback control.
In terms of the physical interpretation discussed in Sec.~\ref{Sec_Stochastic}, the dynamics of system $X$ is determined by the external parameter.  In the presence of feedback control, the time evolution of the  external parameter is determined by $m_1$.
The joint probability distribution of all the variables is then given by
\begin{align}
p(x_N, \cdots, x_1, m_1) =p(x_N |x_{N-1}, m_1) \cdots p(x_2|x_1, m_1) p(m_1|x_1)p(x_1).
\label{feedback}
\end{align}

The Bayesian network corresponding to the above dynamics is characterized as follows.
Let $\mathcal{X} := \{ x_1, \dots, x_N \}$ be the set of the states of measured system $X$, $\mathcal{C} := \{ m_1\}$ be the memory state, and $\mathcal{V} := \mathcal{X} \cup \mathcal{C} = \{ x_1, \dots, x_N, m_1 \}$ be the set of all notes. 
 The causal structure described by Eq.~(\ref{feedback}) is given by  ${\rm pa} (x_k) = \{ x_{k-1}, m_1\}$ for $k\geq 2$, ${\rm pa} (m_1) = \{ x_{1} \}$, and ${\rm pa} (x_1) = \emptyset$ (see also Fig. \ref{example123}).

Since $\mathcal{B}_{k+1} := {\rm pa} (x_{k+1}) \setminus \{ x_k\} = \{ m_1 \}$ for $k \geq 1$, the entropy production (\ref{detailedFT2}) in the heat bath from time $k$ to $k+1$ is given by
\begin{equation}
\Delta s^{\rm bath}_k =  \ln \frac{p(x_{k+1}|x_k, m_1)}{p_B (x_k | x_{k+1} , m_1)}.
\end{equation}
Considering the foregoing argument that $p(x_{k+1}|x_k, m_1)$ depends on $m_1$ through  external parameter $\lambda_k$, we can identify $\Delta s^{\rm bath}_k $ as the heat such that $\Delta s^{\rm bath}_k = -\beta Q_k$.
The total entropy production (\ref{EntropyProduction}) from time $1$ to $N$ is  given by
\begin{align}
\sigma &= \ln \left[ \frac{p(x_1) }{p(x_N)} \prod_{k=1}^{N-1}   \frac{p(x_{k+1}|x_k, m_1)}{p_B (x_k | x_{k+1} , m_1)}\right].
\end{align}

From ${\rm pa} (x_1) = \emptyset$, $\mathcal{C}' =\mathcal{C} = \{ m_1 \}$, and ${\rm pa}_{\mathcal X} (m_1) = \{ x_1 \}$,  we have $I_{\rm fin} =I (x_N: m_1)$, $I_{\rm ini} = 0$, $I_{\rm tr}^1 =I (x_1: m_1)$, and therefore
\begin{equation}
\Theta = I (x_N: m_1) - I (x_1: m_1),
\end{equation}
which is the difference between the initial and the final mutual information. Therefore, the generalized second law~(\ref{GSL}) reduces to
\begin{align}
\langle \sigma \rangle \geq \langle I(x_N: m_1) \rangle -\langle I(x_1:m_1) \rangle.
\label{feedbackGSL}
\end{align}
We note that inequality~(\ref{feedbackGSL}) is equivalent to the generalized second law obtained in Refs.~\cite{Sagawa2,Sagawa3}.

Several simple models that achieves the equality in inequality~(\ref{feedbackGSL}) have been proposed~\cite{Sagawa2012,HorowitzParro,AbreuSeifert}.
In general, the equality in inequality~(\ref{feedbackGSL}) is achieved if and only if a kind of reversibility with feedback control is satisfied~\cite{HorowitzParro}; the reversibility condition is given by
\begin{align}
\prod_{k=1}^{N-1}p(x_{k+1}|x_k , m_1) \cdot p(x_1, m_1) =\prod_{k=1}^{N-1} p_B (x_{k}|x_{k+1}, m_1) \cdot p(x_N, m_1).
\label{feedback_reversibility}
\end{align}
The left-hand side of Eq.~(\ref{feedback_reversibility}) represents the probability of the forward trajectory with feedback control.  The physical meaning of the right-hand side is as follows.  Suppose that we start a backward process just after a forward process by keeping $m_1$ for each trajectory.  In the backward process, we use outcome $m_1$ obtained in the forward process in order to determine the external parameter; we do not perform feedback control in the backward process.  The probability distribution of the backward trajectories is then given by the right-hand side of Eq.~(\ref{feedback_reversibility}).

We next consider a special case that the initial and final states of system $X$ are in thermal equilibrium.
The initial distribution is given by 
\begin{equation}
p(x_1) = p_{\rm eq} (x_1) := e^{\beta (F(1) - E(x_1; 1) )},
\label{canonical_1}
\end{equation}
where $F(1)$ is the initial free energy and $E(x_1;1)$ is the initial Hamiltonian.
Since the final Hamiltonian may depend on outcome $m_1$ due to the feedback control, the final distribution under the condition of $m_1$ is the conditional canonical distribution 
\begin{equation}
p(x_N|m_1) =p_{\rm eq} (x_N|m_1) := e^{\beta (F(m_1) - E(x_1; m_1) )}.
\label{canonical_N}
\end{equation}
Here, $F(m_1)$ is the final free energy and $E(x_1; m_1)$ is the final Hamiltonian, both of which may depend on outcome $m_1$.
 The generalized second law~(\ref{feedbackGSL}) is then equivalent to
\begin{align}
\beta \langle W - \Delta F \rangle &\geq - \langle I(x_1 :m_1) \rangle,
\label{WorkfreeGSL}
\end{align}
where  $W$ is the work and $\Delta F := F(m_1) - F(1)$ is  the free-energy difference.
We note that the ensemble average is needed for $\Delta F$, because $F(m_1)$ is a stochastic quantity due to the stochasticity of $m_1$.
Inequality~(\ref{WorkfreeGSL}) has been derived in~Ref.~\cite{Sagawa1}.
The derivation of inequality~(\ref{WorkfreeGSL}) from (\ref{feedbackGSL}) is as follows.
We first note that
\begin{equation}
\begin{split}
s(x_N) &:= -\ln p(x_N) \\
&= - \ln \left[ p(x_N | m_1) \frac{p(m_1)}{p(x_N | m_1)} \right] \\
&= s(x_N | m_1) + I(x_N : m_1). 
\end{split}
\end{equation}
From Eqs.~(\ref{canonical_1}) and (\ref{canonical_N}), we have
\begin{equation}
s(x_1) = - \beta (F(1) - E(x_1; 1) ), \ s(x_N|m_1) = - \beta (F(m_1) - E(x_1; m_1) ).
\end{equation}
Therefore, we obtain
\begin{equation}
s(x_N) - s(x_1) = -\beta ( \Delta F - \Delta E ) + I(x_N : m_1).
\label{s_I_canonical}
\end{equation}
By substituting the ensemble average of Eq.~(\ref{s_I_canonical}) to inequality~(\ref{feedbackGSL}), we obtain
\begin{equation}
-\beta \langle \Delta F - \Delta E \rangle  + \langle I(x_N : m_1) \rangle - \langle Q \rangle \geq \langle I(x_N: m_1) \rangle -\langle I(x_1:m_1) \rangle.
\label{WorkfreeGSL0}
\end{equation}
By noting the first law $\Delta E = W+Q$, we find that inequality~(\ref{WorkfreeGSL0})  is equivalent to inequality~(\ref{WorkfreeGSL}).

The simplest example of the present setup is the Szilard engine discussed in Sec.~\ref{Sec_Basic} (see also Fig.~\ref{Szilard}).
In this case, the measurement is error-free and the outcome is $m_1 =L$ or $R$ with probability $1/2$, and therefore $\langle I(x_1: m_1) \rangle = \ln 2$.  The final state is no longer correlated with $m_1$ such that $\langle I(x_N: m_1) \rangle = 0$.
The extracted work is $-\langle W \rangle = \beta^{-1}\ln 2$, and the free-energy change is $\langle \Delta F \rangle = 0$.  Therefore, for the Szilard engine, the both-hand sides of inequality (\ref{WorkfreeGSL}) is given by $\ln 2$, and the equality in  (\ref{WorkfreeGSL}) is achieved.  In this sense, the Szilard engine is an optimal information-thermodynamic engine.


\subsection{Example 3: Repeated feedback control with multiple measurements\label{Sec_E_Repeated}}

We consider the case of multiple measurements and feedback control.  
Let $x_k$ be the state of  system $X$ at time $k$ ($=1, \dots, N$).
Suppose that the measurement outcome obtained at time $k$ ($=1, \dots, N-1$), written as $m_k$, is affected by  past trajectory $(x_1, x_2, \cdots, x_k)$ of system $X$.
In other words, the measurement at time $k$ is performed on trajectory $(x_1, x_2, \cdots, x_k)$.
Moreover, we assume that outcome $m_k$ is also affected by sequence $(m_1, \cdots, m_{k-1})$ of the past measurement outcomes, which describes the situation that the way of measuring  $X$ is changed depending on  the past measurement outcomes; such a measurement is called adaptive.
The conditional  probability of $m_k$ is then given by $p(m_k|x_1, \cdots, x_{k-1}, x_k, m_1 \cdots, m_{k-1})$.

Next,   outcome $m_k$ is used for feedback control after time $k$, and the transition probability from $x_k$ to $x_{k+1}$ is written as $p(x_{k+1} | x_k, m_1, \cdots, m_{k-1}, m_k)$.
In this case, we assume that external parameter $\lambda_k$ at time $k$ is determined by memory states $(m_1, \cdots, m_{k-1}, m_k)$.
The joint probability distribution of all the variables is then given by
\begin{equation}
\begin{split}
&{}p(x_1, \cdots, x_N, m_1, \cdots, m_{N-1}	) \\
&= \prod_{k=1}^{N-1} p(x_{k+1}|x_k , m_1, \dots, m_k) p(m_k|x_1, \cdots, x_k, m_1, \cdots, m_{k-1}) \cdot p(x_1).
\end{split}
\label{repeatedFC}
\end{equation}

If outcome $m_k$ is affected only by $x_k$ such that 
\begin{equation}
p(m_k|x_1, \cdots, x_k, m_1, \cdots, m_{k-1}) = p(m_k|x_k),
\end{equation}
the measurement is Markovian and non-adaptive.
If the transition probability from $x_k$ to $x_{k+1}$ depends only on $m_k$ such that 
\begin{equation}
p(x_{k+1} | x_k, m_1,  \cdots, m_{k-1},m_k) = p(x_{k+1} | x_k, m_k),
\end{equation} 
the feedback control is called Markovian.  On the other hand, if  $p(x_{k+1} | x_k, m_1,  \cdots, m_{k-1},m_k)$ depends on $m_l$ with $l<k$, the feedback control is called  non-Markovian, which describes the effect of time-delay of the feedback loop.

The Bayesian network corresponding to the above dynamics is as follows.
Let $\mathcal{X} := \{ x_1, \cdots, x_N \}$, $\mathcal{C} := \{ m_1, \cdots, m_N \}$, and $\mathcal{V} := \mathcal{X} \cup \mathcal{C}$.
The causal structure is characterized by ${\rm pa} (x_k) = \{ x_{k-1}, m_1, \dots, m_{k-1}\}$ for $k\geq 2$, ${\rm pa} (x_1) = \emptyset$, ${\rm pa} (m_k) = \{ x_1, \cdots, x_{k-1}, x_k, m_1, \cdots, m_k  \}$ for $k \geq 2$, and ${\rm pa} (m_1) = \{ x_1 \}$. 
 Figure~\ref{example123} describes the Bayesian network of a special case that 
\begin{equation}
p(m_k|x_k,  \cdots, x_1, m_1, \cdots, m_{k-1}) = p(m_k|x_k,  m_{k-1})
\end{equation}
and ${\rm pa} (m_k) = \{ x_k,  m_{k-1} \}$ for $k \geq 2$.

Since $\mathcal{B}_{k+1} = \{ m_1, \cdots, m_k \}$, the entropy change (\ref{detailedFT2}) in the heat bath from time $k$ to $k+1$ is given by
\begin{equation}
\Delta s^{\rm bath}_k = \ln \frac{p(x_{k+1}|x_k, m_1, \dots, m_k)}{p_B (x_k | x_{k+1} , m_1, \dots, m_k)}.
\end{equation}
If we assume that $p(x_{k+1}|x_k, m_1, \cdots, m_k)$  depends on $(m_1, \cdots, m_k)$ only through external parameter $\lambda_k$, the entropy change is identified with the heat: $\Delta s^{\rm bath}_k = - \beta Q_k$.
The total entropy production (\ref{EntropyProduction}) from time $1$ to $N$ is defined as
\begin{align}
\sigma = \ln \left[ \frac{p(x_1) }{p(x_N)} \prod_{k=1}^{N-1}   \frac{p(x_{k+1}|x_k, m_1, \dots, m_k)}{p_B (x_k | x_{k+1} , m_1, \dots, m_k)}\right].
\end{align}

From ${\rm pa} (x_1) = \emptyset$, $\mathcal{C}' =\mathcal{C} = \{ m_1, \dots, m_{N-1} \}$, and ${\rm pa}_{\mathcal X} (m_k) = \{ x_1, \cdots, x_k \}$, we have $I_{\rm ini} = 0$, $I_{\rm fin} =I (x_N:  ( m_1, \dots, m_{N-1} ) )$, $ I_k^{\rm tr} =I((x_1, \cdots, x_k): m_k | m_1, \dots, m_{k-1})$, and therefore
\begin{equation}
\Theta =  I(x_N: ( m_1, \dots, m_{N-1} ) )  - \sum_{l=1}^{N-1}  I((x_1, \cdots, x_k): m_k | m_1, \dots, m_{k-1}).
\label{repeated_net_information}
\end{equation}
Therefore, the generalized second law~(\ref{GSL}) reduces to
\begin{align}
\langle \sigma \rangle \geq \langle I(x_N:  ( m_1, \dots, m_{N-1} ))  \rangle- \sum_{k=1}^{N-1} \langle I((x_1, \cdots, x_k): m_k | m_1, \dots, m_{k-1}) \rangle.
\label{INEQ}
\end{align}
We note that, in the special case illustrated in Fig.~(\ref{example123}), we have ${\rm pa}_{\mathcal X} (m_k) = \{ x_k \}$ and $ I_k^{\rm tr} =I(x_k: m_k | m_1, \dots, m_{k-1})$.  Therefore, $\Theta$ in Eq.~(\ref{repeated_net_information}) reduces to
\begin{equation}
\Theta =  I(x_N: ( m_1, \dots, m_{N-1} ) )  - \sum_{k=1}^{N-1}  I(x_k: m_k | m_1, \dots, m_{k-1}).
\end{equation}

The equality in Eq. (\ref{INEQ}) holds if and only if the feedback reversibility is satisfied~\cite{HorowitzParro}:
\begin{equation}
\begin{split}
&\prod_{k=1}^{N-1} p(x_{k+1}|x_k , m_1, \dots, m_k) p(m_k|x_1, \cdots, x_k, m_1, \cdots, m_{k-1}) \cdot p(x_1)\\
&= \prod_{k=1}^{N-1} p_B (x_{k}|x_{k+1} , m_1, \dots, m_k)\cdot p(x_N, m_1, \dots, m_{N-1}).
\end{split}
\label{feedback_reversible2}
\end{equation}
The right-hand side of Eq.~(\ref{feedback_reversible2}) represents the probability distribution of the  backward trajectories. In a backward process,   any feedback control is not performed, and the external parameter is changed by using the measurement outcomes obtained in the corresponding forward process.

If the initial and final states of system $X$ are in thermal equilibrium such that $p(x_1) =p_{\rm eq} (x_1) := e^{\beta (F(m_1) - E(x_1; m_1) )}$ and $p(x_N|m_1, \cdots, m_N) =p_{\rm eq} (x_N|m_1, \cdots, m_{N}) := e^{\beta (F(m_1, \cdots, m_{N}) - E(x_1; m_1, \cdots, m_{N}) )}$, inequality (\ref{INEQ}) reduces to, from a similar argument of the derivation of inequality~(\ref{WorkfreeGSL}),
\begin{align}
\beta \langle W - \Delta F \rangle &\geq - \sum_{k=1}^{N-1} \langle I((x_1, \cdots, x_k): m_k | m_1, \dots, m_{k-1}) \rangle,
\end{align}
which has been obtained in Refs.~\cite{HorowitzVaikun,Sagawa2012} for the case of  non-adaptive measurements.

\subsection{Example 4: Markovian information exchanges\label{Sec_E_Bipartite}}

\begin{figure}[htbp]
 \begin{center}
  \includegraphics[width=140mm]{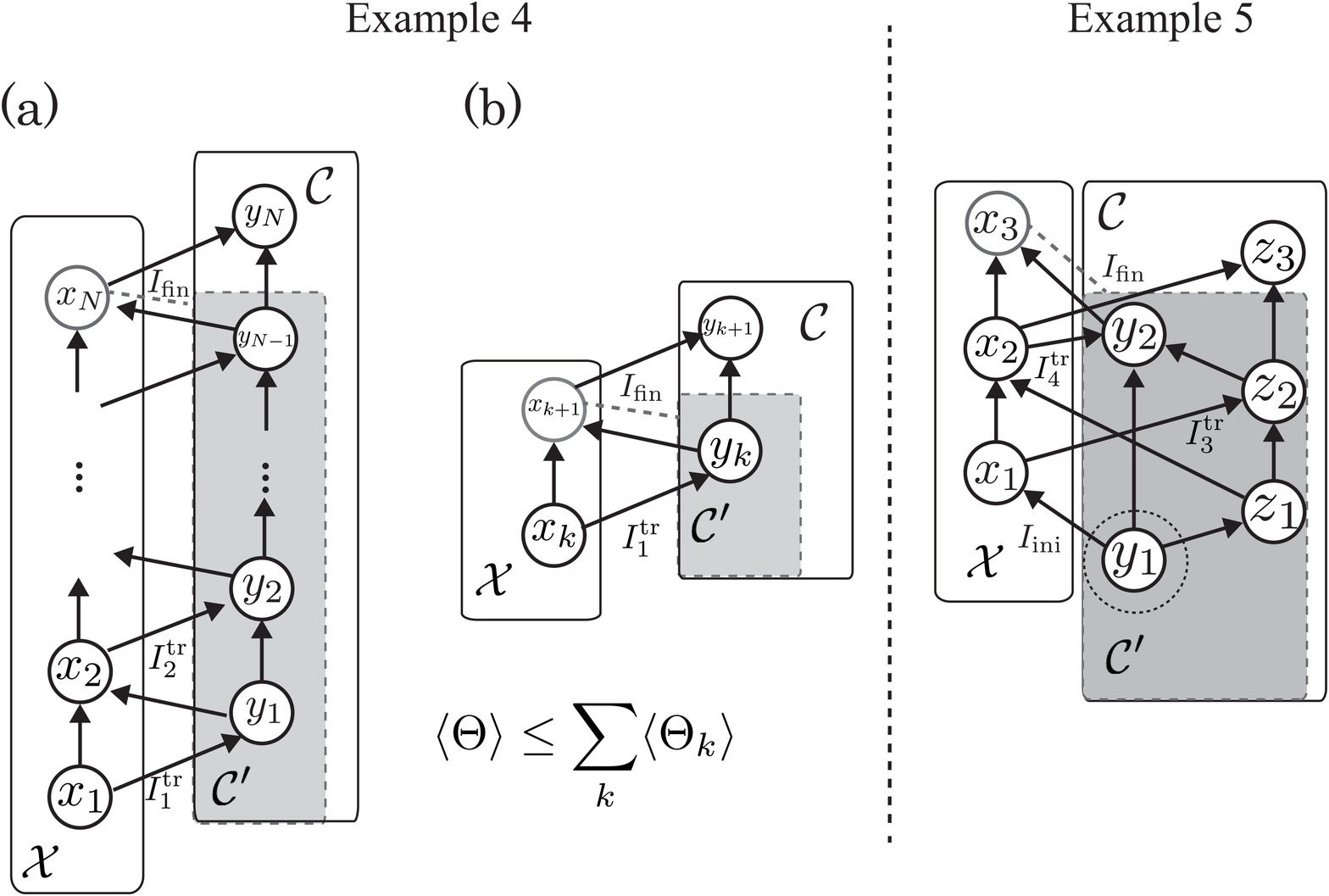}
 \end{center}
 \caption{Bayesian networks of the examples. Example 4: Markovian information exchanges between two systems. (a) Entire dynamics.  (b) A single transition. Example 5: Complex dynamics of three interacting systems.} 
 \label{example45}
\end{figure}

We consider information exchanges between two interacting systems $X$ and $Y$. 
Let $x_k$ and $y_k$ be the states of system $X$ and $Y$ in time ordering $k=1, \dots, N$. 
Suppose that the transition from $x_k$ to $x_{k+1}$ is affected by $y_k$, and the transition from $y_k$ to $y_{k+1}$ is affected by $x_k$ (see also Fig.~\ref{example45} (a)).
This assumption implies that the interaction of $X$ and $Y$ is Markovian.
During the dynamics, the transfer entropy from $X$ to $Y$ and vice versa can be positive, and the mutual information between two systems can change.  Therefore, such dynamics can describe Markovian information exchanges.
In the continuous-time limit, such dynamics are called  Markov jump processes of bipartite systems~\cite{HartichSeifert,HorowitzEsposito}. 
We note that ``bipartite systems" do not mean bipartite graphs in the terminology of Bayesian networks.


The joint probability distribution of all the variables is given by
\begin{align}
p(x_1, y_1, \dots, x_N, y_N) = \prod_{k=1}^{N-1} p(y_{k+1}|x_{k+1}, y_k) p(x_{k+1}|x_k, y_k) \cdot p(y_1|x_1) p(x_1).
\label{multistep}
\end{align}
The transition probability of each step  from $(x_k, y_k)$ to $(x_{k+1}, y_{k+1})$ is given by
\begin{equation}
p(x_{k+1}, y_{k+1}| x_k, y_k) = p(y_{k+1}|x_{k+1}, y_k) p(x_{k+1}|x_k, y_k),
\end{equation}
and correspondingly, the joint probability of $(x_k, y_k, x_{k+1}, y_{k+1})$ is given by
\begin{equation}
p(x_{k+1}, y_{k+1}, x_k, y_k) = p(y_{k+1}|x_{k+1}, y_k) p(x_{k+1}|x_k, y_k)p(y_k|x_k) p(x_k). 
\label{singlestep}
\end{equation}

First, we apply our general argument in Sec.~\ref{Sec_Bayesian} to the entire dynamics~(\ref{multistep}) illustrated in Fig.~\ref{example45} (a).
Let $\mathcal{X} = \{x_1, \dots, x_N\}$ be the set of the states of  $X$, $\mathcal{C} = \{ y_1, \dots, y_N \}$ be the set of the states of $Y$, and $\mathcal{V} := \mathcal{X}  \cup \mathcal{C}= \{x_1, y_1, \dots, x_N, y_N\}$ be the set of all states. The causal structure described by Eq.~(\ref{multistep}) is given by  ${\rm pa} (x_k) = \{x_{k-1}, y_{k-1} \}$ for $k >2$, ${\rm pa} (y_k) = \{x_k, y_{k-1} \}$ for $k>2$, ${\rm pa} (y_1) =\{ x_1 \}$ and ${\rm pa } (x_1) = \emptyset$. 

Since $\mathcal{B}_{k+1} = \{ y_{k} \}$, the entropy change (\ref{detailedFT2}) in the heat bath from time $k$ to $k+1$ is given by
\begin{equation}
\Delta s^{\rm bath}_k = \ln \frac{p(x_{k+1}|x_k, y_{k})}{p_B (x_k | x_{k+1} , y_{k})}.
\label{entropyproductionint}
\end{equation}
The entropy production (\ref{EntropyProduction}) from time $1$ to $N$ is then given by
\begin{align}
\sigma = \ln \left[ \frac{p(x_1) }{p(x_N)} \prod_{k=1}^{N-1}   \frac{p(x_{k+1}|x_k, y_k)}{p_B (x_k | x_{k+1} , y_k)}\right].
\label{entropyproductionint2}
\end{align}

From ${\rm pa} (x_1) = \emptyset$, $\mathcal{C}' ={\rm an}(x_N) \cap \mathcal{C} = \{ y_1, \dots, y_{N-1} \}$, and ${\rm pa}_{\mathcal X} (y_k) = \{ x_k \}$, we have $I_{\rm ini} = 0$, $I_{\rm fin} =I (x_N:  ( y_1, \dots, y_{N-1} ) )$,  $ I_k^{\rm tr} =I(x_k: y_k | y_1, \dots, y_{k-1})$, and therefore
\begin{equation}
\Theta =  I (x_N:  ( y_1, \dots, y_{N-1} ) )  - \sum_{l=1}^{N-1}  I(x_k: y_k | y_1, \dots, y_{k-1}).
\end{equation}
The generalized second law~(\ref{GSL}) then reduces to
\begin{align}
\langle \sigma \rangle \geq \langle \Theta \rangle =\langle I (x_N:  ( y_1, \dots, y_{N-1} ) )   \rangle- \sum_{k=1}^{N-1} \langle I(x_k: y_k | y_1, \dots, y_{k-1}) \rangle.
\label{bound1}
\end{align}

Next, we apply our general argument in Sec.~\ref{Sec_Bayesian} only to a single transition described by Eq.~(\ref{singlestep}), which is illustrated in Fig.~\ref{example45} (b).
 Let $\mathcal{X} = \{x_k, x_{k+1} \}$ be the set of the states of $X$, $\mathcal{C} = \{ y_k, y_{k+1} \}$ be the set of the states of $Y$, and $\mathcal{V} := \mathcal{X}  \cup \mathcal{C}= \{x_k, y_k,  x_{k+1}, y_{k+1} \}$ be the set of all states. 
The causal structure described by Eq.  (\ref{singlestep}) is given by  ${\rm pa} (x_{k+1}) = \{x_{k}, y_{k} \}$, ${\rm pa} (y_{k+1}) = \{x_{k+1}, y_{k} \}$, ${\rm pa} (y_k) =\{ x_k \}$, and ${\rm pa } (x_k) = \emptyset$. 

Since $\mathcal{B}_{k+1} = \{ y_{k} \}$, the entropy change (\ref{detailedFT2}) in the heat bath from time $k$ to $k+1$ is equal to Eq. (\ref{entropyproductionint}). The entropy production of the single transition, written as $\sigma_k$, is given by
\begin{align}
\sigma_k = \ln \left[ \frac{p(x_k) }{p(x_{k+1})} \frac{p(x_{k+1}|x_k, y_k)}{p_B (x_k | x_{k+1} , y_k)}\right].
\end{align}
Here, the sum $\sum_{k=1}^N \sigma_k$ is equal to the entire entropy production $\sigma$ given in Eq.~(\ref{entropyproductionint2}).

From ${\rm pa} (x_k) = \emptyset$, $\mathcal{C}' ={\rm an}(x_{k+1}) \cap \mathcal{C} = \{ y_k \}$, and ${\rm pa}_{\mathcal X} (y_{k+1}) = \{ x_k \}$, we have $I_{\rm ini} = 0$, $I_{\rm fin} =I (x_{k+1}: y_k )$, and $ I_k^{\rm tr} =I(x_k: y_k)$. Denoting $\Theta$ for the single transition by  $\Theta_k$, we obtain
\begin{equation}
\Theta_k =  I (x_{k+1}:y_{k} )  - I (x_{k}:y_{k} ).
\end{equation}
Therefore, the generalized second law~(\ref{GSL}) reduces to
\begin{align}
\langle \sigma_k \rangle \geq  \langle \Theta_k \rangle =\langle I (x_{k+1}:y_{k} )   \rangle-  \langle  I (x_{k}:y_{k} ) \rangle.
\label{INEQb}
\end{align}
By summing up inequality~(\ref{INEQb}) for $k=1,2, \cdots, N-1$, we obtain 
\begin{align}
\langle \sigma \rangle &\geq \langle \Theta^{\rm d} \rangle,
\label{bound2}
\end{align}
where
\begin{equation}
\Theta^{\rm d}  :=  \sum_{k=1}^{N-1} \Theta_k.
\end{equation}
Inequality~(\ref{bound2}) gives another bound of the entire entropy production $\langle \sigma \rangle$. An informational quantity $\langle \Theta^{\rm d} \rangle$  is called the dynamic information flow, which has been studied for the bipartite Markovian jump processes and coupled Langevin dynamics~\cite{Allahverdyan,HartichSeifert,HorowitzEsposito,HorowitzSandberg,Naoto,Shiraish2}.

To summarize the foregoing argument, we have shown two inequalities (\ref{bound1}) and (\ref{bound2}) for the same dynamics described in Fig.~\ref{example45} (a).  Inequality~ (\ref{bound2}) is obtained by summing up inequality~(\ref{INEQb}) for $k=1,2, \cdots, N-1$, where inequality~(\ref{INEQb}) is obtained by applying our general inequality (\ref{GSL}) only to the single transition illustrated in  Fig.~\ref{example45} (b).

We now discuss the relationship of two inequalities  (\ref{bound1}) and (\ref{bound2}).
We can calculate the difference between $\langle \Theta^{\rm d} \rangle$  and $\langle \Theta \rangle$ as
\begin{equation}
\begin{split}
&{} \langle \Theta^{\rm d} \rangle - \langle \Theta \rangle \\
&= \sum_{k=1}^{N-1} \left[ \langle I (x_{k+1}:y_{k} )   \rangle -  \langle I (x_{k}:y_{k} )   \rangle+ \langle I(x_k: y_k | y_1 , \dots, y_{k-1}) \rangle  \right] -\langle I (x_N:  ( y_1, \dots, y_{N-1} ) ) \\
&=\left\langle \ln  \prod_{k=2}^{N-1} \frac{p(x_{k+1}, y_k) p(x_k,  y_1 , \dots, y_k ) }{p(x_{k}, y_{k} ) p(x_{k+1}, y_1 , \dots, y_k ) } \right\rangle \\
&= \sum_{k=2}^{N-1} \left[ \langle I(x_k, (y_1 , \dots, y_{k-1})| y_k) \rangle -\langle I(x_{k+1}, (y_1 , \dots, y_{k-1})| y_k) \rangle  \right] \\
& \geq 0,
\end{split}
\end{equation}
where we used the data processing inequality~\cite{Cover-Thomas}
\begin{align}
\langle I(x_k, (y_1 , \dots, y_{k-1})| y_k) \rangle \geq \langle I(x_{k+1}, (y_1 , \dots, y_{k-1})| y_k) \rangle,
\end{align}
for the following conditional Markov chain:
\begin{equation}
p( x_k, x_{k+1}, y_1, \dots, y_{k-1}| y_k) =p(x_{k+1}| x_k , y_k)p(x_k|y_1, \dots, y_{k-1}, y_k) p(y_1, \dots, y_{k-1}| y_k).
\end{equation}
Therefore, we obtain
\begin{equation}
\langle \sigma \rangle \geq \langle \Theta^{\rm d} \rangle \geq \langle \Theta \rangle,
\end{equation}
which implies that  the dynamic information flow $\langle \Theta^{\rm d} \rangle$ gives a tighter bound of the entire entropy production than  $\langle \Theta \rangle$. 
This hierarchy has been also shown in Ref.~\cite{HorowitzSandberg} for coupled Langevin dynamics.


\subsection{Example 5: Complex dynamics\label{Sec_E_Complex}}

We consider  three systems that interacts  with each other as illustrated in Fig.~\ref{example45}.  In this case, $\mathcal{V} := \{ y_1, x_1, z_1, x_2, z_2, y_2, x_3, z_3 \}$, ${\rm pa} (y_1) = \emptyset$,  ${\rm pa} (x_1) = \{ y_1 \}$,  ${\rm pa} (z_1) =\{ y_1 \}$,  ${\rm pa} (x_2) = \{x_1, z_1 \}$,  ${\rm pa} (z_2) = \{ x_1, z_1 \}$,  ${\rm pa} (y_2) = \{y_1, x_2, z_2\}$,  ${\rm pa} (x_3) = \{ x_2, y_2\}$, and ${\rm pa} (z_3) = \{ x_2, z_2\}$.
The joint probability of $\mathcal{V}$ is given by
\begin{align}
p(\mathcal{V}) =&p(z_3| x_2, z_2) p(x_3|x_2, y_2) p(y_2| y_1, x_2, z_2)  p(z_2|x_1, z_1) p(x_2|x_1, z_1) p(z_1|y_1)p(x_1|y_1) p(y_1).
\end{align}

We focus on system $X$ with $\mathcal{X} := \{ x_1, x_2, x_3\}$.  The other systems are given by $Y$ and $Z$, which constitute $C$ with $\mathcal{C} =  \{ c_1= y_1, c_2= z_1, c_3 = z_2, c_4 =y_2, c_5 =z_3 \}$.
Since $\mathcal{B}_2 = \{z_1 \}$, and $\mathcal{B}_3 = \{y_2 \}$,  the total entropy production  (\ref{EntropyProduction}) is defined as
\begin{align}
\sigma := \ln \frac{p(x_{3} |x_2, y_2)p(x_{2} |x_1, z_1) p(x_1)}{p_B(x_{1} |z_{1}, x_{2}) p_B(x_{2} |y_{2}, x_{3}) p(x_{3})}.
\end{align}

From $\mathcal{C}' =  \{ y_1,  z_1, z_2, y_2 \}$, ${\rm pa} (x_1) = \{ y_1 \}$, ${\rm pa}_{\mathcal X} (y_1) = \emptyset$, ${\rm pa}_{\mathcal X} (z_1) = \emptyset$, ${\rm pa}_{\mathcal X} (z_2) = \{x_1 \}$ and ${\rm pa}_{\mathcal X} (y_2) = \{x_2 \}$, we have $I_{\rm fin} = I(x_3: \{ y_1,  z_1, z_2, y_2 \}) $, $I_{\rm ini} = I(x_1: y_1)$,  $I_{\rm tr}^1 =0$, $I_{\rm tr}^2 = 0$, $I_{\rm tr}^3 = I(x_1: z_2| y_1, z_1)$, and $I_{\rm tr}^4 = I(x_2: y_2| y_1, z_1,z_2)$.
The generalized second law~(\ref{GSL}) then reduces to
\begin{align}
\langle \sigma \rangle \geq&  \langle I(x_3: \{ y_1,  z_1, z_2, y_2 \}) \rangle -\langle  I(x_1: y_1) \rangle-\langle I(x_1: z_2| y_1, z_1)\rangle - \langle I(x_2: y_2| y_1, z_1,z_2) \rangle.
\label{GSLcomp}
\end{align}

\section{Summary and prospects\label{Sec_Summary}}

In this chapter, we have reviewed a general framework of information thermodynamics on the basis of Bayesian networks~\cite{Ito}.
In our framework, Bayesian networks are used to graphically characterize stochastic dynamics of nonequilibrium thermodynamic systems. Each node of a Bayesian network describes a state of a physical system at a particular time, and each edge describes the causal relationship in the stochastic dynamics.  
A simple application of our framework is the setup of ``Maxwell's demon,'' which performs measurements and feedback control, and can extract the work by using information.
Moreover, our framework is not restricted to such simple measurement-feedback situations, but is applicable to a broad class of nonequilibrium dynamics with information exchanges.

Our main result is the generalized second law of thermodynamics (\ref{GSL}).  The entropy production $\langle \sigma \rangle$, which is the sum of the entropy changes in system $X$ and the heat bath, is bounded by an informational quantity $\langle \Theta \rangle$, which consists of the initial and final mutual information between system $X$ and other systems $C$, and the transfer entropy from $X$ to $C$ during the dynamics.  A key ingredient here is the transfer entropy, which quantifies the directional information transfer from a stochastic system to another stochastic system.
The physical meaning of the generalized second law is that the entropy reduction of system $X$ is bounded by the available information about $X$ obtained by $C$.
We note that the generalized second law is derived as a consequence of  the nonnegativity of the relative entropy as shown in (\ref{network_relative}), and also as a consequence of the integral fluctuation theorem (\ref{network_IFT}).
We have also discussed the relationship between the generalized second law with the transfer entropy (\ref{bound1})  and that with  the dynamic information flow (\ref{bound2}) in Sec.~\ref{Sec_E_Bipartite}; the latter second law is stronger.
While we have focused on discrete-time dynamics in this chapter, we can also formulate continuous-time dynamics by Bayesian networks, where we assume that edges represent infinitesimal transitions~\cite{Ito,Ito2}.

For the case of quantum systems, the effect of a single quantum measurement and feedback control has been studied, and the generalizations of the  second law and the fluctuation theorem have been derived in the quantum regime~\cite{SagawaUedaq,Jacobs,SagawaUedaq2,Morikuni,Albash,Funo2,Tajima,Sagawaq,Goold}.
However, the generalization of the formulation with Bayesian networks to the quantum regime has been elusive, which is a fundamental open problem.

Potential applications of information thermodynamics beyond the conventional setup of Maxwell's demon can be found in the filed of biophysics.  In fact, there have been several works that analyze the adaptation process of living cells in terms of information thermodynamics~\cite{HartichSeifert2,SartoriHorowitz,Ito2}.  For example, by applying the generalized second law to biological signal transduction of \textit{Escherichia coli} (\textit{E. coli}) chemotaxis,  we found that the robustness of adaptation is quantitatively characterized by the transfer entropy inside a feedback loop of the signal transduction~\cite{Ito2}.  Moreover, it has been found that the \textit{E. coli} chemotaxis is inefficient (dissipative) as a conventional thermodynamic engine, but is efficient as an information-thermodynamic engine.  These results suggest that information thermodynamics is indeed useful to analyze autonomous information processing in biological systems. 

Another  potential application of information thermodynamics would be machine learning, because neural networks perform stochastic information processing on complex networks. 
In fact, there has been an attempt to analyze neural networks in terms of information thermodynamics~\cite{Hayakawa}.
Moreover, information thermodynamics of  neural information processing in brains would also be another fundamental open problem.


\end{document}